\documentclass[preprint,number,12pt]{elsarticle}  

\usepackage{amssymb}
\usepackage{amsmath,amssymb,amsfonts}
\usepackage{amsthm}
\usepackage{dsfont}
\usepackage{graphicx}
\usepackage{subfigure}
\usepackage{amsbsy}
\usepackage{mathrsfs}
\usepackage{lipsum}
\usepackage{color}
\usepackage{lineno,hyperref}
\modulolinenumbers[5]

\journal{arXiv}

\newcommand{\Randic}{Randi\'{c} }
\newcommand{\Dir}[1]{D_{#1}}
\newcommand{\Is}{\{1, \ldots, N\}}
\newcommand{\Js}{\{0, \ldots, N-1\}}
\newcommand{\Jsz}{\{1, \ldots, N-1\}} 
\newcommand{\Jsh}{\{0, \ldots, n\}} 
\newcommand{\Jshz}{\{1, \ldots, n\}} 
\renewcommand{\cos}{\mathrm{c}}
\renewcommand{\sin}{\mathrm{s}}

\newtheorem{definition}{\textbf{Definition}}
\newtheorem{lemma}{\textbf{Lemma}}
\newtheorem{theorem}{\textbf{Theorem}}
\newtheorem{proposition}{\textbf{Proposition}}
\newtheorem{remark}{\textbf{Remark}}
\newtheorem{corollary}{\textbf{Corollary}}

\renewcommand{\emptyset}{\varnothing} 
\newcommand{\Rset}[1]{\mathbb{R}^{#1}} 
\newcommand{\Cset}[1]{\mathbb{C}^{#1}} 
\newcommand{\Zset}[1]{\mathbb{Z}^{#1}} 
\newcommand{\Nset}[1]{\mathbb{N}^{#1}} 

\def \ii {\boldsymbol{i}} 
\renewcommand{\exp}[1]{e^{#1}} 

\def \w {\mathbf{w}} 
\def \x {\mathbf{x}}

\def \A {\mathbf{A}}

\def \D {\mathbf{D}}

\def \I {\mathbf{I}}

\def \L {\mathbf{L}}

\def \R {\mathbf{R}} 
 
\def \T {\mathbf{T}}

\def\Cmc{\mathcal{C}}

\def\Emc{\mathcal{E}}

\def\Gmc{\mathcal{G}}

\def\Nmc{\mathcal{N}}

\def\Pmc{\mathcal{P}}

\def\Vmc{\mathcal{V}}

\newcommand{\BDelta}{\boldsymbol{\Delta}}

\newtheorem{conj}{Conjecture}

\DeclareMathOperator{\diag}{diag} 
\newcommand{\floor}[1]{\lfloor #1 \rfloor} 
\newcommand{\ceil}[1]{\lceil #1 \rceil} 

\def \normlap		 	{\boldsymbol{\mathcal{L}}} 

\begin{document}

\begin{frontmatter}

\title{On the Characterization of Regular Ring Lattices\\ and their Relation with the Dirichlet Kernel} 

\author{Marco Fabris\fnref{myfootnote}}
\address{Dept. of Information Eng., Univ. of Padova, via Gradenigo 6/B, Padua, 35131, Italy.}
\fntext[myfootnote]{Correspondence: marco.fabris.1@unipd.it}

%
%

\begin{abstract}
Regular ring lattices (RRLs) are defined as peculiar undirected circulant graphs constructed from a cycle graph, wherein each node is connected to pairs of neighbors that are spaced progressively in terms of vertex degree. This kind of network topology is extensively adopted in several graph-based distributed scalable protocols and their spectral properties often play a central role in the determination of convergence rates for such algorithms.
In this work, basic properties of RRL graphs and the eigenvalues of the corresponding Laplacian and \Randic matrices are investigated. A deep characterization for the spectra of these matrices is given and their relation with the Dirichlet kernel is illustrated. Consequently, the Fiedler value of such a network topology is found analytically. With regard to RRLs, properties on the bounds for the spectral radius of the Laplacian matrix and the essential spectral radius of the \Randic matrix are also provided, proposing interesting conjectures on the latter quantities.
\end{abstract}


\end{frontmatter}


\section{Introduction} \label{sec:intro}

\textit{Regular Ring Lattices} (RRLs) are often exploited in a wide range of research fields and they are also known in literature as \textit{$k$-cycles} or \textquotedblleft \textit{pristine worlds}\textquotedblright \cite{BarahonaPecora2002,WuBarahonaTan2011,McKeeArumugam2013,HelleSimonyi2016}. A RRL can be considered a peculiar undirected circulant network \cite{Gray2005} constructed from a cycle graph, wherein each node is connected to pairs of neighbors spaced progressively in terms of vertex degree. 
Remarkably, RRLs are employed in many graph-based distributed scalable algorithms (see, e.g., \cite{MakhdoumiOzdaglar2017,Spielman2007,LovisariZampieri2012,FranceschelliGasparriGiua2013,FabrisMichielettoCenedese2019,FabrisMichielettoCenedese2020,FabrisMichielettoCenedese2022}), as their symmetry can be exploited for design purposes. 
Possible applications for this class of networks may encompass intelligent surveillance of public spaces \cite{LiuLiuMuhammad2018}, tracking-by-detection \cite{HenriquesCaseiroMartins2012}, identification of sparse reciprocal graphical models \cite{AlpagoZorziFerrante2018}, definition of \emph{shift} in graph signal processing \cite{OrtegaFrossardKovacevic2018}, modeling of quantum walks \cite{SevesoBenedettiParis2019}, video circulant sampling schemes \cite{ShuAhuja2011}, compressive three-dimensional sensing techniques \cite{AntholzerWolfSandbichler2019} and sensor network monitoring algorithms \cite{GastparVetterli2005}. The latter examples, in fact, represent only few state-of-the-art topics that motivate this study.
Also, although being of straightforward derivation, a rigorous characterization for the basic and spectral properties of RRLs is lacking or, in some dissertations, incorrect information about their features is provided (see, e.g., the computation of the largest Laplacian eigenvalue $\lambda_{M}$ associated to a RRL in the recently published \cite{GancioRubido2022}).

In light of this premise, RRLs are here examined in detail. 
In particular, the main contributions of this note consist in: 
\begin{enumerate}
	\item the investigation of some of their basic properties;
	\item the spectral analysis of the associated Laplacian and \Randic matrices.
\end{enumerate}
Furthermore, an exact relationship for the spectra of these matrices is yielded through the Dirichlet kernel. A special effort is then directed towards the analytical computation of the Fiedler value \cite{Fiedler1973,Cheeger1969,LiGuoShiu2013}, representing the \textit{algebraic connectivity} of such graphs. With regard to RRLs, properties on the bounds for the spectral radius of the Laplacian matrix \cite{HuiqingMeiFeng2004,Shi2007} and the essential spectral radius of the \Randic matrix \cite{CarliFagnaniSperanzon2008,BrinonSchenato2013} are also provided. Lastly, conjectures on the latter quantities are also proposed.

The remainder of the note is organized as follows. The mathematical preliminaries in Sec. \ref{sec:math_prelim} offer an overview on RRLs. The main results of this work are then presented in Sec. \ref{sec:main_results}, where basic and spectral properties of RRLs are widely explored. The study continues with the discussion in Sec. \ref{sec:discuss_conj}, in which two conjectures related to the spectral radius (for the Laplacian matrix) and the essential spectral radius (for the \Randic matrix) of a RRL are given. Finally, conclusions in Sec. \ref{sec:concl_future_dirs} summarize all the reported findings. 

\paragraph{\textbf{Notation}}
The sets of integer, natural, real, complex numbers are indicated by $\Nset{}$, $\Zset{}$, $\Rset{}$, $\Cset{}$, respectively; whereas, the empty set and the imaginary unit are denoted by $\emptyset$ and $\ii$, respectively. The cosine and sine functions of $\alpha \in \Rset{}$ are respectively denoted with $\mathrm{cos}(\alpha)$ and $\mathrm{sin}(\alpha)$, or abbreviated as $\cos_{\alpha}$ and $\sin_{\alpha}$. The inverse sine and cosine function of $\alpha \in [-1,1]$ are denoted by $\arcsin(\alpha)$ and $\arccos(\alpha)$; while, the inverse tangent function of $\alpha \in \Rset{}$ is denoted by $\arctan(\alpha)$. The complex exponential, floor and ceiling functions are defined respectively as $\exp{} : z \in \Cset{} \mapsto \exp{z} \in\Cset{}  \setminus \{0\}$, $\floor{} : x  \in \Rset{}  \mapsto \floor{x} \in \Zset{}$ and $\ceil{} : x  \in \Rset{}  \mapsto \ceil{x} \in \Zset{}$. Given $N \in \Nset{}  \setminus \{0\}$, the quantity $\theta = \pi / N$ is assigned and used throughout the note to shortly address the $N$-th part of a straight angle $\pi$; moreover, $n = \floor{N/2}$ is set. The modulo and transpose operations are denoted by $\mathrm{mod}$ and $\top$, respectively. Given an $n$-dimensional real-valued vector $\w = (w_{k}) \in \Rset{n}$, the $j$-th \textit{cyclic permutation} over $\w = \begin{bmatrix}
	w_{1} & w_{2} & \cdots & w_{N}
\end{bmatrix}^{\top}$, with $j\in \Zset{}$, is defined as $\w^{j} = \begin{bmatrix}
	w_{1+ (j \mod N)} & w_{1+ (j-1 \mod N)} & \cdots & w_{1+ (j-1+N \mod N)}
\end{bmatrix}^{\top}$ and it holds $\w^{j} = \w$ for all $j \in \Zset{}$ such that $j \mod N = 0$. Also, $\left\|\w\right\|_{1}$ denotes the $1$-norm of $\w$.
Given an $N\times N$-dimensional squared real-valued matrix $\T = (t_{h,k}) \in \Rset{N\times N}$ its $h$-th row is denoted by $\mathrm{row}_{h}(\T)$; furthermore, its $j$-th eigenvalue of is denoted by $\lambda^{\T}_{j}$, with $j \in \Js$. The \textit{spectrum} of $\T$ is defined as the set $\Lambda(\T) = \{\lambda^{\T}_{0}, \ldots, \lambda^{\T}_{N-1} \}$. Notably, it is assumed that eigenvalues $\lambda^{\T}_{j}$ are not necessarily ordered according to their index $j$. 
To conclude, $\I_{N}$ denotes the identity matrix of dimension $N$ and the matrix $\diag(\delta_{1},\ldots,\delta_{N}) \in \Rset{N \times N}$ is equivalent to a squared diagonal matrix $\BDelta = (\delta_{h,k}) \in \Rset{N \times N}$ such that $\delta_{k,k} = \delta_{k}$, for $k \in \Is$; $\delta_{h,k} = 0$, if $h \neq k$.


\section{Preliminaries}\label{sec:math_prelim}
This research begins by briefly illustrating some bases of graph theory and a few mathematical preliminaries about circulant matrices, showing well-known algebraic relations. Also, the definition and a few properties of the Dirichlet kernel are reported.

\subsection{Basic notions of graph theory}\label{ssec:basicgraphtheory}
An \textit{undirected graph} $\Gmc = (\Vmc,\Emc)$ is a networked structure formed by a \textit{vertex set} $\Vmc = \{v_{1},\ldots,v_{N}\}$ and an \textit{edge set} $\Emc \subseteq \Vmc \times \Vmc$, in which each edge $e_{h,k} = (v_{h},v_{k}) = (v_{k},v_{h})$, with $h \neq k$, belongs to $\Emc$ if and only if there exists a connection between vertices $v_{h}$ and $v_{k}$. The cardinality of the edge set is denoted respectively by $M(\Gmc)= \vert \Emc \vert$. Equivalently, the whole structure of $\Gmc$ can be described by the so-called \textit{adjacency matrix} $\A = (a_{h,k}) \in \{0,1\}^{N \times N}$, where $a_{h,k}=1$ if $e_{h,k} \in \Emc$; $a_{h,k}=0$, otherwise. The $k$-th \textit{neighborhood} of vertex $v_{k}$ is then defined as $\Nmc_{k} = \{v_{k} \in \Vmc \;\vert\;  e_{h,k} \in \Emc\}$ and its cardinality $d_{k} = \vert \Nmc_{k} \vert $ is called \textit{vertex degree}. The latter quantity also contributes to the definition of the \textit{degree matrix} $\D = \diag(d_{1},\ldots,d_{N})$. Graph $\Gmc$ is said to be \textit{regular} if all the vertex degrees are equal to some \textit{common degree} $d(\Gmc) \in \Nset{}$. The volume of $\Gmc$ is defined as $\mathrm{vol}(\Gmc) = \sum_{v_{k}\in \Vmc}d_{k}$. Vertex $v_{k}$ is said to be \textit{isolated} if $d_{k}=0$. 
From the above entities, three very relevant matrices associated to $\Gmc$ can be finally defined: the \textit{Laplacian} matrix $\L = \D-\A$ and, assuming that none of the vertices in $\Vmc$ is isolated, the \textit{normalized Laplacian} matrix $\normlap = \D^{-\frac{1}{2}}\L\D^{-\frac{1}{2}}$ and the \textit{\Randic matrix} $\R = \D^{-\frac{1}{2}}\A\D^{-\frac{1}{2}}$~\cite{chung1997spectral,ZhangDong2011,KleinRandic1993,BanerjeeMehatari2016,Rojo2007,Sorgun2013}. Assuming that regularity holds for $\Gmc$, the adjacency, Randi\'{c}, normalized Laplacian and Laplacian matrices associated to $\Gmc$ can be mutually computed through
\begin{equation}\label{eq:matrices_relationsL}
	\L = d(\Gmc)\I_{N}-\A = d(\Gmc)(\I_{N}-\R) = d(\Gmc) \normlap. 
\end{equation}

In addition, a sequence of edges without repetition $\pi_{h,k} \subseteq \Emc$ that links vertices $v_{h}$ and $v_{k}$, in which all traversed vertices are distinct, is called \textit{path}. A \textit{cycle} $\pi_{k}$ passing through vertex $v_{k}$ can be identified as a particular nondegenerate path for which $v_{h}=v_{k}$, i.e. $\pi_{k} = \pi_{k,k}$, with $\pi_{k,k} \neq \emptyset$. If it holds $\pi_{h,k} \neq \emptyset$ for all the couples of vertices $v_{h}$ and $v_{k}$ such that $v_{h}\neq v_{k}$ then $\Gmc$ is said to be \textit{connected}.
The \textit{length of a path} $\pi_{h,k}$ is identified with its cardinality $\vert \pi_{h,k}\vert$, the \textit{distance} between $v_{h}$ and $v_{k}$ is yielded by $\mathrm{dist}(v_{h},v_{k}) = \min \{\vert\pi_{h,k}\vert \;\vert\;  \pi_{h,k}\subseteq \Emc\}$ (note that $\mathrm{dist}(v_{k},v_{k}) = 0$) and the \textit{eccentricity} of vertex $v_{k}$ is computed as $\epsilon (v_{k}) = \max \{\mathrm{dist}(v_{h},v_{k}) \;\vert\;  v_{h} \in \Vmc  \}$. The \textit{diameter} $\phi(\Gmc)$ and \textit{radius} $r(\Gmc)$ of $\Gmc$ are defined as $\phi(\Gmc) = \max\{\epsilon (v_{k}) \;\vert\;  v_{k} \in \Vmc\}$ and $r(\Gmc) = \min\{\epsilon (v_{k}) \;\vert\;  v_{k} \in \Vmc\}$. Also, the \textit{periphery} $\Pmc(\Gmc)$ and \textit{center} $\Cmc(\Gmc)$ of $\Gmc$ are defined as the sets $\Pmc(\Gmc) = \{v_{k} \in \Vmc\; \vert\; \epsilon(v_{k}) = \phi(\Gmc)\}$ and $\Cmc(\Gmc) = \{v_{k} \in \Vmc \; \vert\; \epsilon(v_{k}) = r(\Gmc)\}$. Quantities $g(\Gmc) = \min \{\vert\pi_{k}\vert \;\vert\;  v_{k} \in \Vmc\}$ and $c(\Gmc) = \max \{\vert\pi_{k}\vert \;\vert\;  v_{k} \in \Vmc\}$ are said respectively \textit{girth} and \textit{circumference} of $\Gmc$. 

Lastly, a \textit{cycle graph} $C_{N}$ is an undirected connected regular graph with $N$ vertices such that $d(C_{N}) = 2$; a \textit{complete graph} $K_{N}$ is an undirected connected regular graph with $N$ vertices such that $d(K_{N}) = N-1$; an \textit{edgeless graph} $\overline{K}_{N}$ is a nonconnected regular graph with $N$ isolated vertices ($d(\overline{K}_{N}) = 0$). An undirected connected graph is \textit{Eulerian} if and only if every vertex in it has even degree \cite{Euler1741}. An undirected graph is said \textit{Hamiltonian} if it has a cycle passing through each vertex in it. The smallest number of colors needed to color\footnote{Coloring is intended as labeling each vertex with a nonnegative integer such that no two vertices sharing the same edge have the same label.} a graph $\Gmc$ is denoted by the \textit{chromatic number} $\chi(\Gmc)$. A graph $\Gmc$ with $\chi(\Gmc) = 2$ is said \textit{bipartite}. The following lemma concludes this paragraph.

\begin{lemma}[Handshaking lemma \cite{Euler1741}]\label{lem:handshaking}
	For an undirected graph $\Gmc$, the sum of all its degrees equals twice the number of its edges, i.e. $\mathrm{vol}(\Gmc) = 2M(\Gmc)$.
\end{lemma}

\subsection{Circulant matrices}
In this paragraph, a few fundamental facts about circulant matrices are provided\footnote{Only squared real-valued matrices are considered, as this investigation focuses on undirected (unweighted) RRLs.}.
A circulant matrix is a matrix where each row in it is shifted one entry to the right relative to the previous row vector. The following lines provide its formal definition.
\begin{definition}[Circulant matrix \cite{Gray2005}]\label{def:CirculantM}
	Given an arbitrary vector $\w = (w_{k}) \in \Rset{N}$, the matrix $\T \in \Rset{N\times N}$ is circulant if its $h$-th rows satisfies $\mathrm{row}_{h}(\T) = (\w^{h-1})^{\top}$, for all $h \in \Is$. The vector $\w$ is called \textnormal{generator} of $\T$.
\end{definition}

A circulant topology is thus a structure such that each element in it shares the same ``local panorama'' w.r.t. the other elements. Remarkably, a general expression for the spectrum of circulant matrices can be found. The latter is given in the next theorem.
\begin{theorem}[Spectrum of circulant matrices \cite{Gray2005}]\label{thm:spcircM}
	Let $\T \in \Rset{N \times N}$ be a circulant matrix according to Def. \ref{def:CirculantM}. The spectrum $\Lambda(\T)$ is composed by the eigenvalues $\lambda^{\T}_{j}$ such that
	\begin{equation}\label{eq:eigofcirculants}
		\lambda^{\T}_{j} = \sum\limits_{k=0}^{N-1} w_{k+1} e^{-\ii jk\frac{ 2 \pi }{N}}, \quad \forall j \in \Js.
	\end{equation}
\end{theorem}

\subsection{Definition and properties of the Dirichlet kernel}

According to \cite{BruncknerBruncknerThomson1997}, the definition and few fundamental properties of the Dirichlet kernel are provided in the sequel.

\begin{definition}[Dirichlet kernel \cite{BruncknerBruncknerThomson1997}]\label{def:diri_ker}
	The \textit{Dirichlet kernel} of order $m\in \Nset{}$ is defined as the function $ \Dir{m}: x \in \Rset{} \mapsto \Dir{m}(x)=  \frac{1}{2} \sum_{k = -m}^{m} \exp{\ii k x}$.
\end{definition}

\begin{theorem}[Well-known properties of the Dirichlet kernel \cite{BruncknerBruncknerThomson1997,Wiggins2007,Kirkwood2018}]\label{thm:properties_diri_ker}
	The following properties for the Dirichlet kernel in Def. \ref{def:diri_ker} hold.
	\begin{enumerate}
		\item Each $D_{m}(x)$ is a real-valued, continuous, $2\pi$-periodic, even function and (for $m>0$) assumes both positive and negative values.
		\item For each $m\in \Nset{}$, the Dirichlet kernel can be rewritten as
		\begin{equation}\label{eq:Diri_real_expr}
			D_{m}(x) = \begin{cases}
				\dfrac{\mathrm{sin}\left(\left(m+\frac{1}{2} \right)x\right) }{2\;\!\mathrm{sin}\left(\frac{x}{2} \right)}, \quad &\text{if } x \neq 2\pi \ell, \text{ with } \ell\in \Zset{};\\
				m+\frac{1}{2}, \quad & \text{otherwise;} 
			\end{cases}
		\end{equation}
		or as
		\begin{equation}\label{eq:Diri_real_exprcos}
			D_{m}(x) = \dfrac{1}{2} + \sum\limits_{k=1}^{m} \mathrm{cos}(kx).
		\end{equation}
		\item For each $m\in \Nset{}$ it holds that $\vert D_{m}(x) \vert \leq m+1/2$, $\forall x \in \Rset{}$.
		\item For each $m\in \Nset{} \setminus \{0\}$ the Dirichlet kernel restricted to $[0,2\pi)$ has $2m$ zeros at $x^{\star}_{k} = 2k\pi /(2m+1)$, $\forall k \in \{1,\ldots,2m\}$. In particular, between each pair of consecutive zeros $(x^{\star}_{k},x^{\star}_{k+1})$, $D_{m}(x)$ has one local extremum: a minimum, if $k$ is odd, or a maximum, if $k$ is even.
		\item For each $m\in \Nset{} \setminus \{0\}$ the Dirichlet kernel restricted to $[0,2\pi)$ has one global maximum at $\overline{x}_{0}=0$, for which $D_{m}(\overline{x}_{0})=m+1/2$, and two global minima at $\underline{x}_{1} \in (x^{\star}_{1},x^{\star}_{2})$ and $\underline{x}_{m} = 2\pi-\underline{x}_{1} \in (x^{\star}_{2m-1},x^{\star}_{2m})$. The value of $\underline{x}_{1}$ is approximately given by $\underline{x}_{1} \approx \upsilon x_{1}^{\star}/\pi$, with $\upsilon = 4.493409457909064$.
	\end{enumerate}
\end{theorem}


\section{Main results related to RRLs}\label{sec:main_results}
In this section, the main results on the spectral properties of RRLs are given.
In detail, the RRLs are firstly defined and some basic properties are presented. Then, a spectral analysis of the graph Laplacian matrix $\L$ via the Dirichlet kernel is carried out. This discussion will yield a characterization of its spectrum $\Lambda(\L)$, with particular attention directed towards the \textit{Fiedler value} (i.e. the smallest nonzero eigenvalue of $\L$) and its \textit{spectral radius} (i.e. the largest eigenvalue of $\L$). Then, the investigation continues with a study on the so-called \textit{essential spectral radius} of the \Randic matrix $\R$ associated to a RRL.

\subsection{Definition and basic properties}\label{Ssec:def_DiriGraphs}
Hereafter, a particular kind of circulant graphs is addressed. 
The elements belonging to the class in question are referred to as \textit{RRLs} and described in the following definition.

\begin{definition}[RRL $C_{N}^{m}$]\label{def:Dirichlet_graph}
	Let $N$ and $m$ be two natural numbers such that $N \geq 4$ and $1 \leq m < n=\floor{N/2}$. A RRL $C_{N}^{m} = C_{N}^{m}(\Vmc,\Emc)$ of order $m$ is an undirected graph with $N$ vertices having a circulant adjacency matrix $\A$ generated by a vector $\w \in \{0,1\}^{N}$ whose components are such that 
	\begin{equation}\label{eq:wk_mc}
		w_{k} = \begin{cases}
			1, \quad & \text{if } k \in \{2,\ldots,m+1\} \cup \{N-m+1,\ldots,N\}; \\
			0, \quad & \text{otherwise.}
		\end{cases}
	\end{equation}
\end{definition}

\begin{figure}[b!]
	\centering
	\subfigure[$C_{9}^{1}$]{\includegraphics[width=0.3\columnwidth]{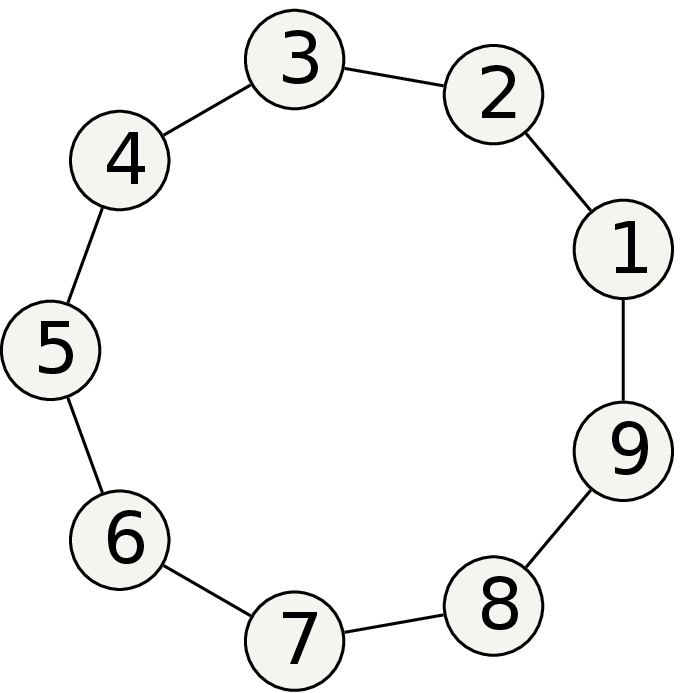}\label{fig:k-ring_graphs_n9k1}}\hspace{0.5cm}
	\subfigure[$C_{9}^{2}$]{\includegraphics[width=0.3\columnwidth]{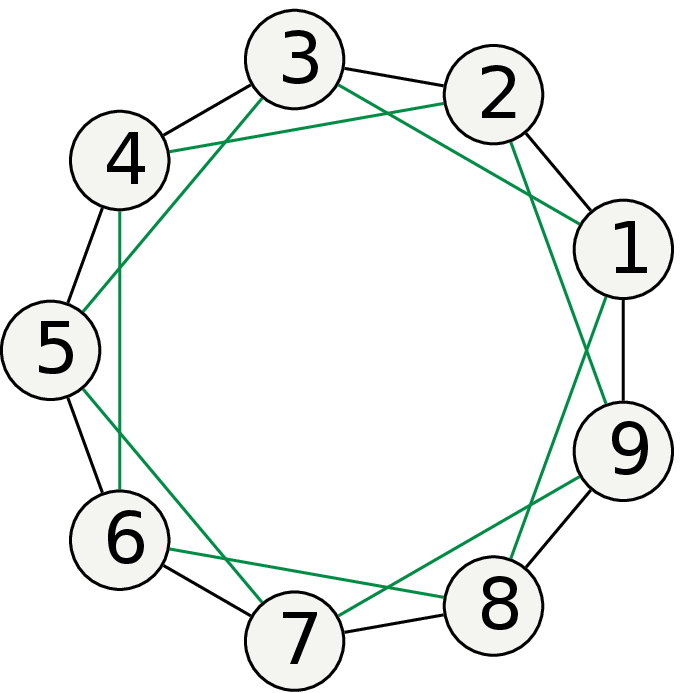}\label{fig:k-ring_graphs_n9k2}}\hspace{0.5cm}
	\subfigure[$C_{9}^{3}$]{\includegraphics[width=0.3\columnwidth]{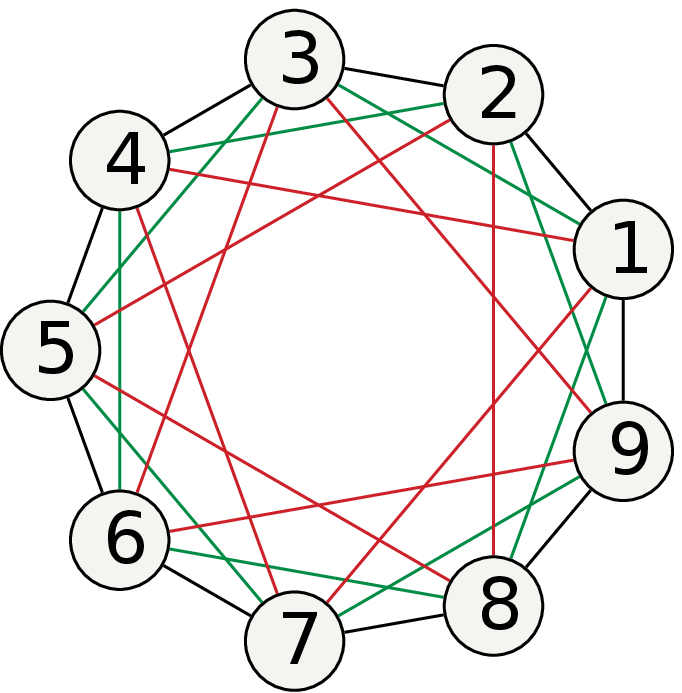}\label{fig:k-ring_graphs_n9k3}}
	\caption{All the three RRLs with $N=9$ vertices. A layer of edges is added for each increasing value of $m \in \{1,2,3\}$: (a) first layer in black, (b) second layer in green, (c) third layer in red.} 
	\label{fig:Dirichlet_graphs}		
\end{figure}

\begin{remark}\label{rmk:order_meaning}
	The order $m$ of a RRL $C_{N}^{m}$ can be interpreted as the identical local field-of-view width of each vertex. In other words, a RRL $C_{N}^{m}$ can ba also said to be a $k$-cycle with $N$ vertices, wherein $k=2m$ neighbors are adjacent to each vertex as depicted in Fig. \ref{fig:Dirichlet_graphs}. 
\end{remark}

It is worth to notice that a RRL $C_{N}^{m}$ is uniquely determined by its number of vertices $N$ and order $m$ only. The following propositions yield all the remaining derived quantities and properties introduced in Ssec. \ref{ssec:basicgraphtheory}.

\begin{proposition}[Regularity and common degree of RRLs]\label{prop:degree}
	Any RRL $C_{N}^{m}(\Vmc,\Emc)$ is regular, with common degree
	\begin{equation}\label{eq:common_deg}
		d(C_{N}^{m}) = 2m.
	\end{equation}
	Consequently, any $C_{N}^{m}$ is Eulerian.
\end{proposition}
\begin{proof}
	The adjacency matrix $\A$ of $C_{N}^{m}$ is circulant and generated by vector $\w$, thus the regularity is shown by observing that for all $v_{k} \in \Vmc$ it holds that $d_{k} = \vert \Nmc_{k}\vert = \left\|\mathrm{row}_{k}(\A) \right\|_{1} = \left\| \w \right\|_{1} = d(C_{N}^{m})$. From \eqref{eq:wk_mc}, the common degree $d(C_{N}^{m})$ is given by the cardinality of $\{2,\ldots,m+1\} \cup \{N-m+1,\ldots,N\}$. Therefore, one has $d(C_{N}^{m}) = (m+1-2+1)+(N-N+m-1+1) = 2m$.
\end{proof} 

\begin{proposition}[Connectivity of RRLs]\label{prop:connectivity}
	Any RRL\\ $C_{N}^{m}(\Vmc,\Emc)$ is connected.
\end{proposition}
\begin{proof}
	By definition, the adjacency matrix $\A = (a_{h,k})$ of $C_{N}^{m}$ satisfies $a_{h,h+1}=1$ for all $h \in \Jsz$. Hence, the path $\pi_{1,N} = \{e_{1,2},e_{2,3}, \ldots, e_{N-1,N}\}$ exists in $C_{N}^{m}$, implying its connectivity.
\end{proof}

\begin{remark}\label{rmk:mcyclelimitcases}
	From Prop. \ref{prop:degree} and Prop. \ref{prop:connectivity} it follows that $C_{N}^{1} = C_{N}$, since RRLs are connected and $d(C_{N}^{m})=2$ if $m=1$. This implies that cycle graphs are a subclass of RRLs and represent a proper basic case in this setting. Moreover, one can also observe that $\lim_{m\rightarrow n} C_{N}^{m} = K_{N}$ follows directly from \eqref{eq:wk_mc}. Therefore, complete graphs represent a degenerate upper limit case for RRLs. One the other hand, one has that $\lim_{m\rightarrow 0} C_{N}^{m} = \overline{K}_{N}$ follows directly from \eqref{eq:wk_mc}. Hence, edgeless graphs represent a degenerate lower limit case for RRLs.
\end{remark}

\begin{corollary}[Volume and number of edges of a RRL]\label{cor:volume_noofedges}
	The volume $\mathrm{vol}(C_{N}^{m})$ and number of edges $M(C_{N}^{m})$ of a RRL $C_{N}^{m}(\Vmc,\Emc)$ are yielded by
	\begin{align}
		\mathrm{vol}(C_{N}^{m}) &= 2mN; \label{eq:volume} \\
		M(C_{N}^{m}) &= mN. \label{eq:noofedges}
	\end{align}
\end{corollary}
\begin{proof}
	By leveraging the definition of volume of a graph and the regularity of RRLs shown in \eqref{eq:common_deg}, relation \eqref{eq:volume} is verified. Whereas, exploiting Lem. \ref{lem:handshaking} on $C_{N}^{m}$, the result in \eqref{eq:noofedges} follows.
\end{proof}

\begin{proposition}[Chromatic number of RRLs]\label{prop:chromatic}
	A RRL $C_{N}^{m}(\Vmc,\Emc)$ has chromatic number
	\begin{equation}\label{eq:chromatic}
		\chi(C_{N}^{m}) = m+1 +(N \mod (m+1)).
	\end{equation}
\end{proposition}
\begin{proof}
	A RRL $C_{N}^{m}$ can be minimally colored exploiting its circulant symmetry. Starting e.g. from vertex $v_{1}$, one can use a group of $m+1$ distinct colors to label subsequent subsets of $m+1$ vertices. In this way, vertices $v_{k}$ share the same color $(k \mod (m+1)) \in \{0,\ldots,m\}$ for all $k$ such that $ 1 \leq k \leq N - (N \mod (m+1))$. Finally, the remaining $(N \mod (m+1))$ vertices need to be labeled with $(N \mod (m+1))$ additional distinct colors.
\end{proof}

\begin{corollary}[Bipartiteness of RRLs]\label{cor:bipartiteness}
	A RRL $C_{N}^{m}(\Vmc,\Emc)$ is bipartite if and only if $m=1$ and $N$ is even.
\end{corollary}
\begin{proof}
	From Prop. \ref{prop:chromatic}, expression \eqref{eq:chromatic} yields $\chi(C_{N}^{m})=2$ if and only if $m=1$ and $N \mod 2 = 0$.
\end{proof}

\begin{proposition}[Diameter and radius of a RRL]\label{prop:diam_radius}
	The diameter $\phi(C_{N}^{m})$ and radius $r(C_{N}^{m})$ of a RRL $C_{N}^{m}(\Vmc,\Emc)$ are yielded by
	\begin{equation}\label{eq:diam_radius}
		\phi(C_{N}^{m}) = r(C_{N}^{m}) = \ceil{\floor{N/2}/m} .
	\end{equation}
\end{proposition}
\begin{proof}
	As each vertex in $C_{N}^{m}$ shares the same local perspective and any $C_{N}^{m}$ is connected (see Prop. \ref{prop:connectivity}), the eccentricity of each $v_{k} \in \Vmc$ is given by $\epsilon(v_{k}) = \epsilon_{0}(C_{N}^{m})$, with constant $\epsilon_{0}(C_{N}^{m}) = \ceil{n/m}$.
\end{proof}

\begin{corollary}[Periphery and center of a RRL]\label{cor:peri_center}
	The periphery $\Pmc(C_{N}^{m})$ and center $\Cmc(C_{N}^{m})$ of a RRL $C_{N}^{m}(\Vmc,\Emc)$ are yielded by
	\begin{equation}\label{eq:peri_center}
		\Pmc(C_{N}^{m})=\Cmc(C_{N}^{m}) = \Vmc.
	\end{equation}
\end{corollary}
\begin{proof}
	Relation \eqref{eq:peri_center} derives from \eqref{eq:diam_radius} in Prop. \ref{prop:diam_radius}.
\end{proof}

\begin{proposition}[Circumference and girth of a RRL]\label{prop:circu_girth}
	The circumference $c(C_{N}^{m})$ and the girth $g(C_{N}^{m})$ of a RRL $C_{N}^{m}(\Vmc,\Emc)$ are yielded by
	\begin{align}
		c(C_{N}^{m}) &= N; \label{eq:circu} \\
		g(C_{N}^{m}) &= \ceil{N/m}.  \label{eq:girth}
	\end{align}
	Consequently, any $C_{N}^{m}$ is Hamiltonian. 
\end{proposition}
\begin{proof}
	Relation \eqref{eq:circu} holds trivially, since $C_{N}^{m}$ always encompasses the cycle graph $C_{N}$ (see Rmk. \ref{rmk:mcyclelimitcases}). This implies that any $C_{N}^{m}$ is Hamiltonian. Whereas, \eqref{eq:girth} is retrieved similarly to what done with eccentricity in Prop. \ref{prop:diam_radius}.
\end{proof}

In Tab. \ref{tab:MCproperties}, all the discussed properties of RRLs are summarized. 

\begin{table}[t]
	\small
	\centering
	\begin{tabular}{|c|c|c|}
		\hline
		\multicolumn{1}{|c|}{No. Vertices} & \multicolumn{1}{c|}{No. Edges} & \multicolumn{1}{c|}{Common degree} \\ \hline
		\centering
		$N\!: \;N\geq 4$ & $ M(C_{N}^{m}) = mN $  &  $ d(C_{N}^{m}) = 2m $ \\\hline 
		\hline
		\multicolumn{1}{|c|}{Order} & \multicolumn{1}{c|}{Volume} & \multicolumn{1}{c|}{Chromatic number} \\ \hline
		\centering
		$m\!: \; 1\leq m<\floor{N/2}$ & $ \mathrm{vol}(C_{N}^{m}) = 2mN$  &  $\chi(C_{N}^{m}) = m+1+(N \mod (m+1))$ \\\hline
		\hline
		\multicolumn{1}{|c|}{Diameter} & \multicolumn{1}{c|}{Periphery} & \multicolumn{1}{c|}{$$Circumference$$}\\ \hline
		\centering
		$\phi(C_{N}^{m})=\ceil{\floor{N/2}/m}$     &  $ \Pmc(C_{N}^{m})= \Vmc $     &    $c(C_{N}^{m}) = N$  \\\hline 
		\hline
		\multicolumn{1}{|c|}{Radius} & \multicolumn{1}{c|}{Center} & \multicolumn{1}{c|}{Girth}\\ \hline
		\centering
		$r(C_{N}^{m})=\ceil{\floor{N/2}/m}$     &  $ \Cmc(C_{N}^{m}) = \Vmc $     &    $g(C_{N}^{m}) = \ceil{N/m}$  \\\hline
	\end{tabular}
	\caption{Basic topological quantities of a RRL $C_{N}^{m}(\Vmc,\Emc)$.}
	\label{tab:MCproperties}
\end{table}


\subsection{Spectral analysis} 

The analysis starts by showing the key insight to examine the spectral properties of
RRLs via the theoretical support provided by the properties of the Dirichlet kernel $D_{m}$. A characterization for the eigenvalues of the Laplacian matrix $\L$ associated to the RRLs in terms of $D_{m}$ is given by the following theorem, explaining the reason why $m$ is considered the order for this class of graphs. To avoid heavy notation, $d = d(C_{N}^{m})=2m$ is adopted henceforward.

\begin{theorem}[Spectral characterization of RRLs]\label{thm:spect_char}
Let $\mathbf{L}$ be the graph Laplacian matrix associated to a RRL $C_{N}^{m}$.
Setting $\theta=\pi /N$, the spectrum $\Lambda(\mathbf{L})$ can be expressed in function of the Dirichlet kernel $D_{m}$ as
\begin{align}
	\lambda^{\mathbf{L}}_{j} &= 1+2\left(m-D_{m}(2\theta j)\right), \quad \forall N\geq 4, \forall m \in \{1,\ldots,n-1\} , \label{eq:eigL1} 
\end{align}
with $\lambda^{\mathbf{L}}_{N-j} =  \lambda^{\mathbf{L}}_{j}$, $\forall j \in \Jshz$.
Furthermore, the following properties hold for all $N\geq 4$ and $m \in \{1,\ldots,n-1\}$.
\begin{enumerate}
	\item Each eigenvalue $\lambda^{\mathbf{L}}_{j}$ belongs to $[0,4m]$ for all $j \in \Js$.
	\item Eigenvalue $\lambda_{0}^{\mathbf{L}} = 0$ is simple, i.e. it has algebraic multiplicity $1$.
	\item If $\exists \lambda^{\mathbf{L}}_{j^{\star}} = 4m$ for some $j^{\star}\in \Jshz$ then eigenvalue $\lambda^{\mathbf{L}}_{j^{\star}}$ is simple.
\end{enumerate}
\end{theorem}

\begin{proof}
Let $\A$ be the adjacency matrix of $C_{N}^{m}$ generated by the vector $\w$, according to Def. \ref{def:Dirichlet_graph}. Recalling that given $\alpha \in \Rset{}$ and a matrix $\T \in \Rset{N \times N}$ it holds that $\lambda^{\I_{N}+\alpha \T}_{j} = 1+\alpha \lambda^{\T}_{j}$ for all $j\in \Js$ (see \cite{PetersenPedersen2008}), the relations between the $j$-th eigenvalue of matrices in 
\eqref{eq:matrices_relationsL} are the following:
\begin{equation}\label{eq:lambdaL_ARLn}
	\lambda^{\L}_{j} = d-\lambda^{\A}_{j} = d(1-\lambda^{\R}_{j})= d \lambda^{\normlap}_{j}.
\end{equation}
Now, the $j$-th eigenvalue of the adjacency matrix $\A$ can be computed resorting to \eqref{eq:eigofcirculants} in Thm. \ref{thm:spcircM} and Def. \ref{def:diri_ker} as follows:
\begin{align}
	\lambda_{j}^{\A} &= \sum\limits_{k=0}^{N-1} w_{k+1} e^{-2\ii jk\theta} = \sum\limits_{k=1}^{m} e^{-2\ii jk\theta} + \sum\limits_{k=N-m}^{N-1} e^{-2\ii jk\theta} \nonumber\\
	& = \sum\limits_{k=1}^{m} e^{-2\ii jk\theta} + \sum\limits_{k=1}^{m} e^{2\ii jk\theta} = \left(  \sum\limits_{k=-m}^{m} e^{\ii k(2\theta j)}\right) -1 \nonumber \\
	&= 2 (D_{m}(2\theta j) -1/2), \quad \forall N\geq 4, \forall m \in \{1,\ldots,n-1\}. \label{eq:lambdaA_Dir}
\end{align}
Therefore, substituting \eqref{eq:lambdaA_Dir} in \eqref{eq:lambdaL_ARLn} and leveraging Prop. \ref{prop:degree} and Thm. \ref{thm:properties_diri_ker}, 
relation \eqref{eq:eigL1} can be found. In particular, $\lambda^{\mathbf{L}}_{N-j} =  \lambda^{\mathbf{L}}_{j}$ holds $\forall j \in \Jshz$ since $D_{m}(x)$ is $2\pi$-periodic and even (see Thm. \ref{thm:properties_diri_ker}). 

Lastly, regarding the rest of the statement, authors in \cite{landau1981bounds} have already shown that matrix $\R$ has eigenvalues belonging to the interval $[-1, 1]$, where $\lambda^{\R}_{0} = 1$ and, possibly, $\lambda^{\R}_{j^{\star}} = -1$ for some $j^{\star} \in \Jshz$ are both associated to a single eigenvector. Also, leveraging the connectivity of $C_{N}^{m}$ shown in Prop. \ref{prop:connectivity}, it holds that $\lambda_{0}^{\normlap}=0$ and $0<\lambda_{j}^{\normlap}\leq 2$ for all $j \in \Jsz$ (see \cite{chung1997spectral}). 
Resorting to \eqref{eq:lambdaL_ARLn}, one has
\begin{equation*}
	\lambda_{j}^{\normlap} = 1-m^{-1} (D_{m}(2\theta j) -1/2), \quad \forall N\geq 4, \forall m \in \{1,\ldots,n-1\}
\end{equation*}
and the thesis easily follows.
\end{proof}

The result provided by Theorem \ref{thm:spect_char} contributes with equalities \eqref{eq:eigL1}, yielding an interesting interconnection between the Dirichlet kernel and the eigenvalues of the
graph Laplacian matrix $\L$ corresponding to a RRL. The analysis proceeds by focusing on the extremal (maximum and minimum) eigenvalues belonging to the restricted spectrum $\Lambda_{0}(\L) = \Lambda(\L) \setminus \{\lambda_{0}^{\L}\} \subseteq (0,4m]$. In the following lines, some properties related to the \textit{Fiedler value} $\nu(\L) = \min_{\lambda\in \Lambda_{0}(\L)} \{\lambda\}$ and the \textit{spectral radius} $\rho(\L) = \max_{\lambda \in \Lambda(\L)} \{\lambda\}$ of a RRL Laplacian matrix are provided.

\begin{theorem}[Algebraic connectivity of the RRLs]\label{thm:fiedler_value}
Let $C_{N}^{m}$ be a RRL and $\L$ be the corresponding Laplacian matrix with eigenvalues $\lambda_{j}^{\L}$ given by \eqref{eq:eigL1}. Then the algebraic connectivity of a RRL $C_{N}^{m}$ is yielded by the Fiedler value $\nu(\L)$ of $\L$, whose expression is
\begin{equation}\label{eq:maxeigrel}
	\nu(\L) = \lambda_{1}^{\L} = \lambda_{N-1}^{\L}, \quad \forall N\geq 4, \; \forall m\in \{1,\ldots,n-1\} .
\end{equation}
Moreover, one has $\nu(\L) \in (0,2m]$ and $\nu(\L)=2m$ if and only if $2m=N-2$.
\end{theorem}

\begin{proof}
Exploiting the symmetry of $\Lambda(\L)$ discussed in Thm. \ref{thm:spect_char}, let us restrict w.l.o.g. this analysis to eigenvalues in $\Lambda_{0}(\L)$ indexed by $j \in \Jshz$. It can be noticed that relations \eqref{eq:Diri_real_expr} and \eqref{eq:lambdaL_ARLn} lead to 
\begin{equation}\label{eq:lambdaRj}
	\lambda_{j}^{\R} = m^{-1} (D_{m}(2\theta j) -1/2), \quad \forall N\geq 4, \forall m \in \{1,\ldots,n-1\},
\end{equation}
which can be leveraged to prove that $\lambda^{\mathbf{L}}_{1} < \lambda^{\mathbf{L}}_{j}$ holds for all $j\in \{2,\ldots,n\} $ by verifying the following chain of inequalities: 
\begin{equation}\label{ineq:sinsinsinsin}
	\lambda^{\R}_{1} > \lambda^{\R}_{j}  \;\Longleftrightarrow\;   D_{m}(2\theta) > D_{m}(2\theta j ) \;\Longleftrightarrow\;	\dfrac{\sin_{(2m+1)\theta}}{\sin_{\theta}}  > \dfrac{\sin_{ (2m+1)\theta j}}{\sin_{\theta j}}.
\end{equation}
Considering that $\sin_{z} = z\prod_{k=1}^{+\infty} (1-\frac{z^{2}}{k^{2}\pi^{2}})$, $\forall z\in \Cset{}$ (see formula 4.3.89 in \cite{AbramowitzStegun1972}), the following inequality can be derived from the rightmost expression in \eqref{ineq:sinsinsinsin}:
\begin{equation}\label{ineq:prodsin}
	\prod\limits_{k=1}^{+\infty} \dfrac{k^{2}N^{2} - (\mathrm{2m}+1)^{2}}{k^{2}N^{2}-1}   >    \prod\limits_{k=1}^{+\infty}  \dfrac{k^{2}N^{2} - (\mathrm{2m}+1)^{2}j^{2}}{k^{2}N^{2}-j^{2}}.
\end{equation}
For relation \eqref{ineq:prodsin} to be satisfied, it is sufficient to prove that:\\
(i) the $k$-th factor on the l.h.s. is strictly positive for all $k\in \mathbb{N} \setminus \{0\}$;\\
(ii) the $k$-th factor on the l.h.s. is strictly greater than the $k$-th factor on the r.h.s. for all $k\in \mathbb{N}\setminus \{0\}$.\\
Property (i) is verified, since this requirement boils down to the identity $2m < N-1 \leq kN-1$ for all $k\in \mathbb{N}\setminus \{0\}$; while, property (ii) is also satisfied, as this leads to the identities $m>0$ and $j>1$ for all $k \in \mathbb{N}\setminus\{0\}$. Hence, relation \eqref{eq:maxeigrel} is now proven.

To conclude, it is worth to show that $\lambda^{\R}_{1}$ is nonnegative for any given $C_{N}^{m}$. By \eqref{eq:Diri_real_expr} and \eqref{eq:lambdaRj} one has the relation
\begin{equation}\label{eq:ineq_s_lambdaR1}
	\lambda_{1}^{\R} = m^{-1}(D_{m}(2\theta)-1/2) \geq 0 \;\Longleftrightarrow\; \sin_{(2m+1)\theta} \geq \sin_{\theta} .
\end{equation}
Since $\theta \in (0,\pi/4]$ and $m\geq 1$, the last inequality in \eqref{eq:ineq_s_lambdaR1} holds true for any admissible $(N,m)$. Also, strict equality in \eqref{eq:ineq_s_lambdaR1} is satisfied for $m=n-1$ and even $N$.
Therefore, $\lambda^{\R}_{1}$ belongs to the interval $[0,1)$ and, by \eqref{eq:lambdaL_ARLn}, one has $\lambda^{\mathbf{L}}_{1} \in (0,2m]$ and $\lambda^{\mathbf{L}}_{1} = 2m$ if and only if $2m=N-2$.
\end{proof}

\begin{theorem}[Spectral radius properties of RRLs]\label{thm:spectral_radius}
Let $C_{N}^{m}$ be a RRL and $\L$ be the corresponding Laplacian matrix with eigenvalues $\lambda_{j}^{\L}$ given by \eqref{eq:eigL1}. Also, let $j^{\star}$ be an index for which the spectral radius of $\L$ can be expressed as $\rho(\L) = \lambda_{j^{\star}}^{\L} = \lambda_{N-j^{\star}}^{\L}$. Then the following properties are satisfied for all $N\geq 4$.
\begin{enumerate}
	\item For all $m \in \{1,\ldots,n-1\}$ index $j^{\star}$ is yielded by\footnote{If there exist multiple distinct values $j_{1},j_{2},\ldots$ of $j$ minimizing \eqref{eq:jstarcomputation} then $j^{\star} = \min \{j_{1},j_{2},\ldots\}$ is assumed to be the principal minimizer.}
	\begin{equation}\label{eq:jstarcomputation}
		j^{\star} =  \underset{j \in \{2,\ldots,n\}}{\arg \min} \{D_{m}(2\theta j)\} \in \{2,\ldots,n\}.
	\end{equation}\label{point:jstarcomputation}
	In particular, the below partial characterization for $j^{\star}$ can be given.
	\begin{enumerate}
		\item If $m=1$ then $j^{\star} = n$. \label{point:ln_1L_j=n/2} 
		\item Let $b_{2} =  \arccos(-1/4)/(2\theta) $. If $m=2$ then $j^{\star} \in \{\floor{b_{2}},\ceil{b_{2}}\}$.
		\label{point:ln_1L_j-m=2}
		\item Let $b_{3} =\arccos\left((\sqrt{7}-1)/6\right) /(2\theta) $. \\If $m=3$ then $j^{\star} \in \{\floor{b_{3}},\ceil{b_{3}}\}$.
		\label{point:ln_1L_j-m=3}
		\item Let $b_{4}^{-} = \arccos\left( (6 \mathrm{cos}((4\arctan(1/\sqrt{5})-\pi)/3)-1)/8\right) /(2\theta)$ and $b_{4}^{+} = \arccos\left( (-6 \mathrm{cos}(4\arctan(1/\sqrt{5})/3)-1)/8\right) /(2\theta)$, where $b_{4}^{-} < b_{4}^{+}$. If $m=4$ then $j^{\star} \in \{\floor{b_{4}^{-}},\ceil{b_{4}^{-}},\floor{b_{4}^{+}},\ceil{b_{4}^{+}}\}$.
		\label{point:ln_1L_j-m=4}
		\item Let us assign \\ $b_{5,1} = \sqrt{\sqrt{11}-5~\mathrm{cos}((\arctan(\sqrt{55}/11)+\pi)/3)} $,\\ $b_{5,2} = \sqrt{\sqrt{11}-5~\mathrm{cos}((\arctan(\sqrt{55}/11)-\pi)/3)}$,\\ $b_{5,3} = \sqrt{\sqrt{11}+5~\mathrm{cos}(\arctan(\sqrt{55}/11)/3)}$,\\ 
		$b_{5}^{-} = \arccos((\sqrt[\leftroot{-1}\uproot{1}4]{11} (b_{5,1}+b_{5,2}+b_{5,3})-1)/10)/(2\theta)$ 
		and\\ $b_{5}^{+} = \arccos((\sqrt[\leftroot{-1}\uproot{1}4]{11} (b_{5,1}-b_{5,2}-b_{5,3})-1)/10)/(2\theta)$, where $b_{5}^{-} < b_{5}^{+}$. If $m=5$ then
		$j^{\star} \in \{\floor{b_{5}^{-}},\ceil{b_{5}^{-}},\floor{b_{5}^{+}},\ceil{b_{5}^{+}}\}$.
		\label{point:ln_1L_j-m=5}
		\item If $m=n-1$ then $j^{\star} = 2$. 
		\label{point:ln_1L_j-m=max}
	\end{enumerate}
	\item For all $m \in \{1,\ldots,n-1\}$ it holds that $\rho(\L) \in (2m+1,4m]$, with $\rho(\L) = 4m$ if and only if $N$ is even and $m=1$. \label{point:interval_ln_1L}
	\item For all $m \in \{1,\ldots,n-1\}$ there exists $\underline{j} \in \{2,\ldots,n\}$ such that $\underline{j} \leq j^{\star} \leq n$ is satisfied. Moreover, the expression of $\underline{j}$ is given by \label{point:jjbounds}
	\begin{equation}\label{eq:lower_bound_j_star}
		\underline{j} = 1+\floor{N/(2m+1)}.
	\end{equation} 
\end{enumerate}
\end{theorem}

\begin{proof}
Let us restrict w.l.o.g. the analysis to $j \in \Jshz$ by exploiting the symmetry shown in Thm. \ref{thm:spect_char}. Each property of the statement is proven in the sequel.

\ref{point:jstarcomputation} Expression \eqref{eq:jstarcomputation} holds as it is equivalent to
\begin{equation}\label{eq:jstarcomputation2}
	j^{\star} =  \underset{j \in \{2,\ldots,n\}}{\arg \max}  \{\lambda^{\L}_{j}\}= \underset{j \in \{2,\ldots,n\}}{\arg \max} \{1+2\left(m-D_{m}(2\theta j)\right)\},
\end{equation}
as it directly descends from \eqref{eq:eigL1}. Remarkably, in \eqref{eq:jstarcomputation2}, $j=0$ and $j=1$ are excluded, as $\lambda_{0}^{\L}=0$ and $\lambda_{1}^{\L}=\nu(\L)$ are proven to be the smallest eigenvalues of $\L$ (see Thm. \ref{thm:spect_char} and Thm. \ref{thm:fiedler_value}).

\ref{point:ln_1L_j=n/2}. Setting $m = 1$, equality $\lambda^{\mathbf{L}}_{j} =  4\sin^{2}_{\theta j}$ follows by resorting to the triple angle identity $\sin_{3z} = 3\sin_{z}-4\sin_{z}^{3}$, $\forall z \in \Cset{}$. Hence, for $m=1$, the $j$-th eigenvalue $\lambda^{\mathbf{L}}_{j}$ is trivially maximized by selecting $j^{\star}= \lfloor N/2 \rfloor = n$. Also, note that if $N$ is even then $\rho(\L) = 4\sin^{2}_{\theta n}=4$ holds in accordance to property \ref{point:interval_ln_1L}.

\ref{point:ln_1L_j-m=2}. For $m=2$, the global minimum of the Dirichlet kernel $D_{m}(x)$ is obtained for $x = \underline{x}_{1} = \arccos(-1/4)$ by solving the trigonometric first-degree equation descending from $D^{\prime}_{m}(x) =0$, where $D^{\prime}_{m}(x) = -\sum_{k=1}^{m} k~\!\mathrm{sin}(kx)$ is the derivative w.r.t. $x$ of $D_{m}(x)$ (see \eqref{eq:Diri_real_exprcos}), and verifying that $2\pi/5 = x_{1}^{\star}<\underline{x}_{1}<x_{2}^{\star} = 4\pi /5$. Imposing $2\theta j \approx \underline{x}_{1}$ leads to 
the thesis.

\ref{point:ln_1L_j-m=3}. For $m=3$, the global minimum of the Dirichlet kernel $D_{m}(x)$ is obtained for $x = \underline{x}_{1} = \arccos\left((\sqrt{7}-1)/6\right)$ by solving the trigonometric second-degree equation descending from  $D^{\prime}_{m}(x) =0$ and verifying that $2\pi/7 = x_{1}^{\star}<\underline{x}_{1}<x_{2}^{\star} = 4\pi /7$. Imposing $2\theta j \approx \underline{x}_{1}$ leads to the thesis. However, differently from the previous point, an additional check is here needed. In particular, because of the presence of a second local minimum\footnote{This is actually attained for $j=n$ when $N$ is even, as $2\theta j = \underline{x}_{2}$ holds for $j=N/2$.} $\underline{x}_{2} = \pi$ with ordinate $D_{m}(\underline{x}_{2})=-1/2$, it is sufficient to show that $j_{3}^{\star} \in \{\floor{b_{3}},\ceil{b_{3}}\}$ satisfies $D_{m}(2\theta j_{3}^{\star})\leq -1/2$ in order. 
In this direction, one can find all the values of $x\in (0,\pi]$ such that $D_{m}(x)=-1/2$. These solutions are yielded by $\tilde{x}_{1} = \pi/3$, $\tilde{x}_{2} = \pi/2$ and, obviously, $\tilde{x}_{3} = \underline{x}_{2} = \pi$. To conclude the proof, it is sufficient to demonstrate that $\tilde{x}_{2}-\tilde{x}_{1}\geq2\theta$. This inequality is however verified only if $N\geq 12$. Checking all the instances characterized by $4\leq N \leq 11$ and $m=3$, one has $j^{\star} \neq n$ for $N\neq 8$; and $j^{\star} = j_{2}^{\star} = 2$ or $j^{\star}=4$, for $N=8$. Thus, the thesis follows.

\ref{point:ln_1L_j-m=4}. 
This statement is obtained by solving the trigonometric third-degree equation descending from $D_{m}^{\prime}(x)=0$, similarly to what shown in point \ref{point:ln_1L_j-m=2}.

\ref{point:ln_1L_j-m=5}. 
This statement is obtained by solving the trigonometric fourth-degree equation descending from $D_{m}^{\prime}(x)=0$, similarly to what shown in point \ref{point:ln_1L_j-m=3}.

\ref{point:ln_1L_j-m=max}. It can be easily shown that, for all $j \in \Jshz$, one has $D_{n-1}(2 \theta j) = (-1)^{j+1}/2$, if $N$ is even; $D_{n-1}(2 \theta j) = (-1)^{j+1} \cos_{\theta j}$, if $N$ is odd. Therefore, $j=2$ minimizes $D_{n-1}(2 \theta j)$.

\ref{point:interval_ln_1L}.  
By \eqref{eq:jstarcomputation}, the maximum value for $\lambda_{j}^{\L}$ is attained when $D_{m}(2\theta j)$ is minimized in $j$. So, let us consider $D_{m}(2\theta y)$, with $y\in \Rset{}$. According to Thm. \ref{thm:properties_diri_ker}, the zeros of $D_{m}(2\theta y)$ can be expressed as $y_{k}^{\star} = kN/(2m+1)$ for all $k \in \{1,\ldots,2m\}$. Remarkably, each consecutive interval $(y_{k}^{\star},y_{k+1}^{\star})$ has uniform length $N/(2m+1) > 1$. Since the Dirichlet kernel is negative over intervals $(y_{k}^{\star},y_{k+1}^{\star})$ with odd $k$ and $y_{k+1}^{\star}-y_{k}^{\star} >1$, there exists an integer $j^{\star}$ for which $D_{m}(2\theta j^{\star})$ is negative. As a consequence, it holds that $ (1+2m-\lambda^{\L}_{j^{\star}})/2 = D_{m}(2\theta j^{\star}) < 0 $, implying that $\lambda^{\L}_{j} > 2m+1$.
Moreover, $\rho(\L) = 2d = 4m$ holds if and only if $C_{N}^{m}$ is bipartite \cite{LiuLu2010}, namely when $N$ is even and $m=1$, as shown in Cor. \ref{cor:bipartiteness}.

\ref{point:jjbounds}. Since $\rho(\L) > 2m+1$ follows from $D_{m}(2j^{\star}\theta) < 0$, a lower bound $\underline{j}$ for $j^{\star}$ can be computed by solving $D_{m}(2\theta j) < 0$ for $j \in \{2,\ldots,n\}$. Via \eqref{eq:Diri_real_expr}, this leads to the following system of inequalities
\begin{equation}\label{eq:jineqbounds}
	\begin{cases}
		j < 2\ell N / (2m+1) ;  \\
		j > (2\ell-1) N/ (2m+1) ;
	\end{cases}
\end{equation}
where $\ell \in \Zset{}$. Clearly, the first inequality in \eqref{eq:jineqbounds} requires that $\ell \geq 1$, as $j$ is a positive index. Therefore, to find $\overline{j}$, it is imposed $\ell =1$. Consequently, since $1 < N/(2m+1) < n$ holds true for any admissible values of $(N,m)$, the second inequality in \eqref{eq:jineqbounds} evaluated at $\ell =1$ provides the lower bound \eqref{eq:lower_bound_j_star}.
\end{proof}

\begin{remark}
It is worth to note that index $j^{\star}$ 
can be easily computed in closed-form solutions through $D^{\prime}_{m}(x)=-\sum_{k=1}^{m} k~\!\mathrm{sin}(kx)=0$ for $m\in \{1,2,3,4,5,n-1\}$. However, for $m$ such that $6 \leq m \leq n-2$ this kind of expressions cannot be obtained in such a way, since $D^{\prime}_{m}(x)=0$ leads to trigonometric equations having degree five or higher.
\end{remark}

\subsubsection{Essential spectral radius analysis} 
According to \cite{GarinSchenato2010}, the \emph{essential spectral radius} of a row-stochastic\footnote{The matrix $\T = (t_{h,k}) \in \Rset{N \times N}$ is said \textit{row-stochastic} if all its entries $t_{h,k}$ belong to interval $ [0,1]$ for all $ h,k =1,\ldots,N$ and $\left\|\mathrm{row}_{h}(\T) \right\|_{1} = 1$ for all $h=1,\ldots,N$.} \Randic matrix $\R$ can be defined as
\begin{equation}\label{eq:stochspectrad_def}
\sigma(\R) = \underset{\lambda \in \Lambda_{0}(\R)}{\max} \{\vert \lambda \vert\} ,
\end{equation}
where $\lambda_{0}^{\R}=1$ holds and $\Lambda_{0}(\R) = \Lambda(\R) \setminus \{\lambda_{0}^{\R}\}$ is assigned.
Remarkably, the essential spectral radius of a \Randic matrix $\R$ associated to a RRL $C_{N}^{m}$ complies with definition in \eqref{eq:stochspectrad_def} for all admissible $(N,m)$, since $\R = d^{-1}\A$ is a row-stochastic matrix with eigenvalues $\vert \lambda_{j}^{\R}\vert \leq 1$, $\forall j \in \Js$, and $\lambda_{0}^{\R} = 1$. A study on $\sigma(\R)$ for each $C_{N}^{m}$ is thus reported by starting from the next lemma.

\begin{lemma}\label{lem:-l2>l1}
Let $\R$ be the \Randic matrix of a RRL $C_{N}^{m}$ and $\theta=\pi/N\in (0,\pi/4]$. There exists a real number $m^{\star} \in (0,n)$ such that if $m \geq m^{\star}$ then $\lambda^{\R}_{1}+\lambda^{\R}_{2} \leq 0$, with the equality holding if and only if $m = m^{\star}$. Moreover, the value of $m^{\star}$ is yielded by
\begin{equation}\label{eq:mstar}
m^{\star} = \theta^{-1} \arcsin\left(\sqrt{x^{\star}}\right),
\end{equation}
where $x^{\star}$ is the unique solution belonging to $(0,1)$ of the cubic equation
\begin{equation}\label{eq:kbar_poly}
\mathrm{p}_{\theta}(x) = x^{3} +a_{2} x^{2}  + a_{1} x  +a_{0} = 0,
\end{equation}
in which $a_{2} = -(\cos_{2\theta}+5)/2$, $a_{1} = (4\cos_{2\theta}^{2} + 7\cos_{2\theta} +13)/8$, $a_{0} = -(3\cos_{2\theta} +1)^{2}/16$.
\end{lemma}

\begin{proof}
From \eqref{eq:lambdaRj}, the eigenvalues of the \Randic matrix $\R$ can be rewritten using the prosthaphaeresis formula for the difference of two sines as
\begin{equation}\label{eq:prosteigF}
\lambda^{\R}_{j} = \begin{cases}
	\dfrac{\mathrm{sin}(m\theta j)\;\!\mathrm{cos}((m+1)\theta j)}{m \;\! \mathrm{sin}(\theta j )}, \quad &\text{if } j \in \Jsz ; \\
	1, \quad &\text{if } j=0.
\end{cases} 
\end{equation}
Thus, inequality $\lambda^{\R}_{1} + \lambda^{\R}_{2} \leq 0$ can be written as follows by means of the triple angle identities $\cos_{3z}=4\cos_{z}^{3}-3\cos_{z}$, $\sin_{3z}=3\sin_{z}-4\sin_{z}^{3}$, $\forall z \in \Cset{}$, the Werner's formula for the product of two cosines and the basic trigonometric rules:
\begin{equation}\label{ineq:trigon_kbar}
(1-\cos_{2\theta}^{2}) (5-4\sin_{m\theta}^{2})^{2} \sin_{m \theta}^{2} \geq (1-\sin_{m\theta}^{2})(4\cos_{2\theta}(1-\sin_{m\theta}^{2}) +1-\cos_{2\theta})^{2}.
\end{equation}

Now, assigning $x = \sin_{m\theta}^{2} \in (0,1)$, inequality \eqref{ineq:trigon_kbar} can be solved in $m$ by resorting to equation \eqref{eq:kbar_poly} and determining the solutions of $\mathrm{p}_{\theta}(x) \geq 0$.
The application of the Routh-Hurwitz criterion to $\mathrm{p}_{\theta}(x)$, as illustrated in Table \ref{tab:RHpthetax}, ensures that there exists a solution $x^{\star}$ of $\mathrm{p}_{\theta}(x)$ having a strictly positive real part for any value of $\theta$, since each pair of subsequent terms in the second column exhibits an alternating sign.
\begin{table}[h!]
\centering
\begin{tabular}{c|c|c}
	$x^{3}$ &  $1$ & $(4\cos_{2\theta}^{2} + 7\cos_{2\theta} +13)/8$ \\
	$x^{2}$ & $-(\cos_{2\theta}+5)/2$ & $-(3\cos_{2\theta} +1)^{2}/16$ \\ \hline
	$x^{1}$ & $(2\cos_{2\theta}^{3}+9\cos_{2\theta}^{2}+21\cos_{2\theta}+32)/(4(\cos_{2\theta}+5))$ & $0$ \\ \hline
	$x^{0}$ & $-(3\cos_{2\theta} +1)^{2}/16$ & $0$ \\
\end{tabular}
\caption{Routh array for polynomial $\mathrm{p}_{\theta}(x)$.}
\label{tab:RHpthetax}
\end{table}

Analogously, in order to show that $ x^{\star} $ has real part smaller than $1$ for all $\theta$, the Routh-Hurwitz criterion can be also applied to $-\mathrm{p}_{\theta}(y)$, setting $y = 1-x$. This leads to the analysis reported in Table \ref{tab:RHpthetay}: the fact that each pair of subsequent terms in the second column exhibits an alternating sign finally ensures that $x^{\star} \in (0,1)$, provided that $x^{\star} \in \Rset{}$.
\begin{table}[h!]
\centering
\begin{tabular}{c|c|c}
	$y^{3}$ &  $1$ & $(4\cos_{2\theta}^{2}-\cos_{2\theta} -3)/8$ \\
	$y^{2}$ & $-(1-\cos_{2\theta})/2$ & $-(1-\cos_{2\theta}^{2} )/16$ \\ \hline
	$y^{1}$ & $(2\cos_{2\theta}^{2}-\cos_{2\theta}-2)/4$ & $0$ \\ \hline
	$y^{0}$ & $-(1-\cos_{2\theta}^{2} )/16$ & $0$ \\
\end{tabular}
\caption{Routh array for polynomial $-\mathrm{p}_{\theta}(y)$.}
\label{tab:RHpthetay}
\end{table}

According to method 3.8.2 in \cite{AbramowitzStegun1972}, equation \eqref{eq:kbar_poly} can be solved by setting
\begin{equation}\label{eq:qrtheta}
q_{\theta}  = a_{1}/3-a_{2}^{2}/9, \quad r_{\theta} = (a_{1}a_{2}-3a_{0})/6-a_{2}^{3}/27,
\end{equation}
through the computation and observation of the discriminant
\begin{equation}\label{eq:discr_thirdeqdeg}
\Delta_{\theta} = q_{\theta}^{3} + r_{\theta}^{2} = \dfrac{7\left(1-\cos_{2\theta}\right)   \left(1-\cos_{2\theta}^{2}\right)  (\cos_{2\theta}+13/14) }{1728}  \left(\cos_{2\theta}-\dfrac{1}{2}\right)^{2} \geq 0.
\end{equation}
Expression in \eqref{eq:discr_thirdeqdeg} is strictly positive if and only if factor $(\cos_{2\theta}-1/2)^{2}$ is grater than zero: this occurs for values of $\cos_{2\theta} \neq 1/2$, i.e. for $N\neq 6$. In this case, the presence of only one real solution is guaranteed and it is yielded via \eqref{eq:qrtheta}, \eqref{eq:discr_thirdeqdeg} by
\begin{equation}\label{eq:solution3}
x^{\star} =  - \dfrac{a_{2}}{3}+ \sqrt[\leftroot{-1}\uproot{1}3]{r_{\theta} + \sqrt{\Delta_{\theta}}}  +  \sqrt[\leftroot{-1}\uproot{1}3]{r_{\theta} - \sqrt{\Delta_{\theta}}}  .
\end{equation}
Otherwise, for $N=6$, the discriminant $\Delta_{\theta}$ vanishes and the solutions for \eqref{eq:kbar_poly} are given by $\left\lbrace 1/4, 5/4,5/4\right\rbrace$. In fact, for $N=6$, expression \eqref{eq:solution3} boils down to $x^{\star} = 1/4 \in (0,1)$.

Finally, the thesis in \eqref{eq:mstar} is proven by inverting relation $x^{\star}=\sin_{m^{\star}\theta}^{2}$.
\end{proof}

In conclusion, some theoretical results on the essential spectral radius of $\R$ for RRLs are stated in the next theorem.

\begin{theorem}[Essential spectral radius properties of RRLs]\label{thm:ESR}
Let $C_{N}^{m}$ be a RRL and $\R$ the corresponding \Randic matrix with eigenvalues $\lambda_{j}^{\R}$ given by \eqref{eq:lambdaRj}. Also, according to Thm. \ref{thm:spectral_radius}, let $j^{\star} \in \{2,\ldots,n\}$ be computed as in \eqref{eq:jstarcomputation}. 
Then, for the essential spectral radius $\sigma(\R)$, the following properties are satisfied for all $N\geq 4$.
\begin{enumerate}
\item For all $m\in \{1,\ldots,n-1\}$, it holds that $\sigma(\R) = \max \{\lambda_{1}^{\R},-\lambda_{j^{\star}}^{\R} \}$ or, equivalently, $ \sigma(\R) = \max\{\lambda_{N-1}^{\R},-\lambda_{N-j^{\star}}^{\R}\}$, with $\sigma(\R) \in ((2m)^{-1},1] \subseteq (1/2,1]$. In particular, it holds $\sigma(\R) = \vert \lambda_{\gamma}^{\R}\vert  = \vert \lambda_{N-\gamma}^{\R}\vert $, with $\gamma$ such that
\begin{equation}\label{eq:characterizationofjcirc}
	\gamma = \underset{j \in \Jshz}{\arg \min} \left\lbrace \bigg\vert  D_{m}(2\theta j) -\dfrac{1}{2} \bigg\vert  \right\rbrace  \in \{1,j^{\star}\}.
\end{equation}
\label{point:esr_fundamental}
\item If $m = 1$ then $\sigma(\R) = -\lambda^{\R}_{n} = -\lambda^{\R}_{N-n} $.  \label{point:lambda!jn_2}
\item It holds that $\sigma(\R) = 1$ if and only if $N$ is even and $m = 1$. \label{point:boundslambda!}
\item If $m \geq m^{\star}$, with $m^{\star}$ defined as in Lem. \ref{lem:-l2>l1}, then it holds that $\sigma(\R) = -\lambda_{j^{\star}}^{\R} = -\lambda_{N-j^{\star}}^{\R} \leq -\lambda_{2}^{\R} = -\lambda_{N-2}^{\R}$.  \label{point:lambda!j2} 
\item If $m=n-1$ then $\sigma(\R) = -\lambda_{2}^{\R} = -\lambda_{N-2}^{\R} $. \label{point:lambda!m=n-1}
\end{enumerate}
\end{theorem}

\begin{proof}
By the symmetry of the Dirichlet kernel, eigenvalues of $\R$ in \eqref{eq:lambdaRj} also exhibit the property $\lambda_{j}^{\R} = \lambda_{N-j}^{\R}$, for all $j \in \Jshz$. At the light of this observation, the following analysis is restricted to indexes $j \in \{1,\ldots,n\}$.

\ref{point:esr_fundamental}. Exploiting relation \eqref{eq:lambdaL_ARLn} and the fact that $\lambda_{1}^{\L} \leq 2m$ (see Thm. \ref{thm:fiedler_value}) and $\lambda_{j^{\star}}^{\L} > 2m+1$ (see Thm. \ref{thm:spectral_radius}), it follows that $\lambda_{1}^{\R}\geq 0$ and $\lambda_{j^{\star}}^{\R}< -(2m)^{-1}\leq -1/2$ are the largest eigenvalues of $\R$ in absolute value. In particular, \eqref{eq:characterizationofjcirc} is directly derived from \eqref{eq:lambdaRj} applied to \eqref{eq:stochspectrad_def}.

\ref{point:lambda!jn_2}. Applying \eqref{eq:prosteigF} with $m=1$, it holds that $\lambda_{j}^{\R} = \cos_{2\theta j}$. If $N$ is even then $j=n$ is trivially selected to provide the essential spectral radius $\sigma(\R) = -\lambda_{n}^{\R}=1$. Otherwise, for odd $N$, $j=1$ or $j=n$ can be both selected, since $\sigma(\R) = \lambda_{1}^{\R} = \cos_{2\theta}$ or, equivalently, $\sigma(\R) = -\lambda_{n}^{\R} = -\cos_{2\theta n}=\cos_{2\theta}$.

\ref{point:boundslambda!}. In the previous point it is already shown that $\sigma(\R)=1$ if $m=1$ and $N$ is even. To prove that $\sigma(\R)=1$ also implies that $m=1$ and $N$ is even, property \ref{point:interval_ln_1L} of Thm. \ref{thm:spectral_radius} is invoked. Indeed, recall that $\rho(\L)=4m$ holds if and only if $C_{N}^{m}$ is bipartite, namely it has even $N$ and $m=1$. Relation \eqref{eq:lambdaL_ARLn} is then used to conclude.

\ref{point:lambda!j2}. Lem. \ref{lem:-l2>l1} shows that if $m\geq m^{\star}$ then $\lambda_{1}^{\R}+\lambda_{2}^{\R} \leq 0$. Since, in general, it holds that $\lambda_{1}^{\R}\geq 0$, then, if $m\geq m^{\star}$, one has $\lambda_{2}^{\R} \leq -\lambda_{1}^{\R} \leq0$. In particular, if $m> m^{\star}$ then $\vert \lambda_{2}^{\R} \vert > \lambda_{1}^{\R}$ holds; therefore, $j=1$ cannot be a valid index for an eigenvalue $\lambda_{j}^{\R}$ selected to compute $\sigma(\R)$ in this case. As a consequence, if $m\geq m^{\star}$ then $\sigma(\R) = -\lambda_{j^{\star}}^{\R} \leq -\lambda_{2}^{\R}$.

\ref{point:lambda!m=n-1}. Again, for all $j \in \Jshz$, one has $D_{n-1}(2 \theta j) = (-1)^{j+1}/2$, if $N$ is even; $D_{n-1}(2 \theta j) = (-1)^{j+1} \cos_{\theta j}$, if $N$ is odd. Thus, to prove this statement, it is just required to check that $-\lambda_{2}^{\R} > \lambda_{1}^{\R}$ holds true for all odd $N\geq 5$. The latter inequality leads to an identity;  hence, $\lambda_{2}^{\R}$ is the eigenvalue that satisfies \eqref{eq:stochspectrad_def} if $m=n-1$. 
\end{proof}


\section{Further discussions and numerical examples}\label{sec:discuss_conj}

This section reports a discussion on a couple of conjectures about the spectral radius $\rho(\L)$ of the Laplacian matrix $\L$ and on the essential spectral radius $\sigma(\R)$ of the \Randic matrix $\R$ associated to a RRL $C_{N}^{m}$. Meaningful numerical examples are also brought as evidence for these ideas.

\subsection{Conjecture on a potential upper bound for $j^{\star}$}\label{ssec:conjubstar}

Let us consider the statement of Thm. \ref{thm:spectral_radius}. Finding analytically an upper bound $\overline{j}$ for $j^{\star}$, similarly to what done in \eqref{eq:lower_bound_j_star}, may not be trivial. Nonetheless, an interesting conjecture on this particular bound is here given.

\begin{conj}[An upper bound for $j^{\star}$]\label{conj:upperboundjstar}
Under the same assumptions of Thm. \ref{thm:spectral_radius}, there exists $\overline{j} \in \{2,\ldots,n\}$ such that $j^{\star} \leq \overline{j}$ and its expression is yielded by
\begin{equation}\label{eq:upper_bound_j_star}
\overline{j} = \ceil{3N/(4m+2)-1/2}, \quad \forall N\geq 4, \forall m \in \{1,\ldots,n-1\}.
\end{equation}
\end{conj}

\begin{remark}
Considering $\underline{j}$ and $\overline{j}$ computed respectively as in \eqref{eq:lower_bound_j_star} and \eqref{eq:upper_bound_j_star}, the following properties holding for $N\geq4$ can be easily proven to support the fact that $\overline{j}$ may represent a good candidate upper bound for $j^{\star}$.
\begin{enumerate}
\item If $m\geq \widetilde{m}$, where 
\begin{equation}\label{eq:tildem}
	\widetilde{m} = 3N/10-1/2,
\end{equation}
then one has $\overline{j}=2$.
\item One has $\overline{j}=n$ if and only if $m=1$. This also implies that for $m=1$ expression in \eqref{eq:upper_bound_j_star} is, in fact, a valid upper bound for $j^{\star}$. Moreover, if $m\geq 2$ then $\overline{j} < 2N/(2m+1)  = x_{2}^{\star}/(2\theta)  < n$ (see Thm. \ref{thm:properties_diri_ker})
\item If $m=2$ (and $N\geq 6$) then $\overline{j}$ is, in fact, a valid upper bound for $j^{\star}$, since $\overline{j} = \ceil{(3N-5)/2} \leq 2N/5 =  x_{m}^{\star}/(2\theta)$ (see Thm. \ref{thm:properties_diri_ker}).
\item One has $2 \leq \underline{j} \leq \overline{j} \leq n$, in which $\underline{j} = \overline{j} $ holds if and only if at least one of the following three cases is verified: (i) $ 3N/14-1/2 \leq m \leq  N/4-1/2$; (ii) $m\geq \widetilde{m}$; (iii) $N \mod (2m+1) = 0$ and $m \geq  N/6 -1/2$.
\end{enumerate}
\end{remark}

The upper bound in \eqref{eq:upper_bound_j_star} is figured out after the attempt to minimize $D_{m}(2\theta j)$ w.r.t. $j$. Observing that $\sin_{\theta j}$ is strictly increasing for $j\in \Jshz$, relation \eqref{eq:upper_bound_j_star} is derived by choosing the smallest $j\in \{2,\ldots,n\}$ such that $\vert (2m+1)\theta j-(3\pi/2+2\ell\pi )\vert$, $\ell \in \Zset{}$, be minimum and, to make treatable the latter expression, $\ell = 0$ is forced. The aim of this careful selection is twofold: on one hand, we want to obtain a small positive value for the denominator of $D_{m}(2\theta j)$ and, on the other hand, a large (in modulus) negative value for the numerator of $D_{m}(2\theta j)$, see \eqref{eq:Diri_real_expr}. However, in general, there may exist values of $j> \overline{j}$ that render the numerator of $D_{m}(2\theta j)$ even more negative! 
This consideration is crucial; indeed, the reasoning shown for the derivation of formula (36) in \cite{GancioRubido2022} can be trivially disproved taking for instance $(N,m)=(67,2)$, for which it holds that $j^{\star} = 19$ (there, $j^{\star}=20$ is wrongly claimed).\\
Nevertheless, one has $\overline{j} \geq  \ceil{\upsilon N/(\pi(2m+1))} \approx \ceil{\underline{x}_{1}/(2\theta)} $, as $3/2 > \upsilon/\pi$ (see Thm. \ref{thm:properties_diri_ker}). Also, expression \eqref{eq:upper_bound_j_star} has been tested in simulation for all $N$ such that $4 \leq N \leq 10000$ and any relative admissible value of $m$. Remarkably, no counterexample has been found in any of the tested instances; hence, this fact suggests that $\overline{j}$ in \eqref{eq:upper_bound_j_star} might represent a suitable upper bound for $j^{\star}$.

The following remark illustrates the potential implications of Conj. \ref{conj:upperboundjstar}.

\begin{remark}\label{rmk:ifconj1true}
Let $m^{\star}$ and $\widetilde{m} $ be defined as in \eqref{eq:mstar} and \eqref{eq:tildem}, respectively.
If Conj. \eqref{conj:upperboundjstar} verifies then one would have these further implications.
\begin{enumerate}
\item $\rho(\L) = \lambda_{n}^{\L} = \lambda_{N-n}^{\L}$ holds for all $N\geq 4$ if and only if $m=1$. Thus, property \ref{point:ln_1L_j=n/2} in Thm. \ref{thm:spectral_radius} would be reinforced.
\item With reference to the essential spectral radius $\sigma(\R)$, one has, $\forall N\geq 4$, $\sigma(\R) = -\lambda^{\R}_{n} = -\lambda^{\R}_{N-n} $ if and only if $m=1$. Thus, property \ref{point:lambda!jn_2} in Thm. \ref{thm:ESR} would be reinforced.
\item If $m\geq \widetilde{m}$ then $\rho(\L) = \lambda_{2}^{\L} = \lambda_{N-2}^{\L}$ holds for all $N\geq 4$.\label{point:m2l2}
\item Considering again $\sigma(\R)$, if $m \geq \max \{m^{\star},\widetilde{m}\}$ then it holds that $\sigma(\R) = -\lambda_{2}^{\R} = -\lambda_{N-2}^{\R}$ for all $N\geq 4$. Thus, property \ref{point:lambda!j2} in Thm. \ref{thm:ESR} would be reinforced. \label{point:mstarm2}
\item The search space of minimization in \ref{point:ln_1L_j-m=3} and \ref{point:ln_1L_j-m=4} 
of Thm. \ref{thm:spectral_radius} would be reduced into $j\in \{\floor{b_{4}^{-}},\ceil{b_{4}^{-}}\}$ and $j\in \{\floor{b_{5}^{-}},\ceil{b_{5}^{-}}\}$, respectively.
\item The spectral radius $\rho(\L) $ could be computed efficiently through binary search algorithm, as it can be shown that $D_{m}(2\theta y)$ restricted to $y\in [\underline{j},\overline{j}]$ has one global minimum given by $y = \underline{x}_{1}/(2\theta) \approx \upsilon N/(\pi(2m+1))  $ (see Thm. \ref{thm:properties_diri_ker}). Consequently, the computation of $\sigma(\R) = \max \{1-\nu(\L)/(2m),-1+\rho(\L)/(2m)\}$ would also result more efficient.
\item A direct estimate $\widehat{j}^{\star} \in [\underline{j},\overline{j}]$ for $j^{\star}$ could be provided by averaging $\underline{j}$ and $\overline{j}$ through convex combinations. For instance, given $\alpha \in [0,1]$, one can choose\footnote{For all $N$ and $m$ such that $4\leq N \leq 2000$ and $1\leq m < n$, coefficient $\alpha= 0.1313$ seems a good value to reduce the estimation error $\vert j^{\star}-\widehat{j}^{\star}\vert$, with $\widehat{j}^{\star}$ computed as in \eqref{eq:jstarest}.}
\begin{equation}\label{eq:jstarest}
	\widehat{j}^{\star} = \begin{cases}
		n, \quad &\text{if } m=1; \\
		\ceil{b_{2}^{-}-1/2}, \quad & \text{if } m=2; \\
		\ceil{b_{3}^{-}-1/2}, \quad & \text{if } m=3; \\
		\ceil{b_{4}^{-}-1/2}, \quad & \text{if } m=4; \\
		\ceil{b_{5}^{-}-1/2}, \quad & \text{if } m=5; \\
		2, \quad & \text{if } m=n-1; \\
		\floor{\alpha \underline{j} + (1-\alpha) \overline{j} + 1/2}, \quad &\text{otherwise}.
	\end{cases}
\end{equation}

\end{enumerate}
\end{remark}

\subsection{Numerical examples for $4\leq N \leq 11$}\label{Ssec:numerical_examples}
\begin{figure}[ht!]
	\centering
	\subfigure[$N=4$]{\includegraphics[width=0.49\columnwidth]{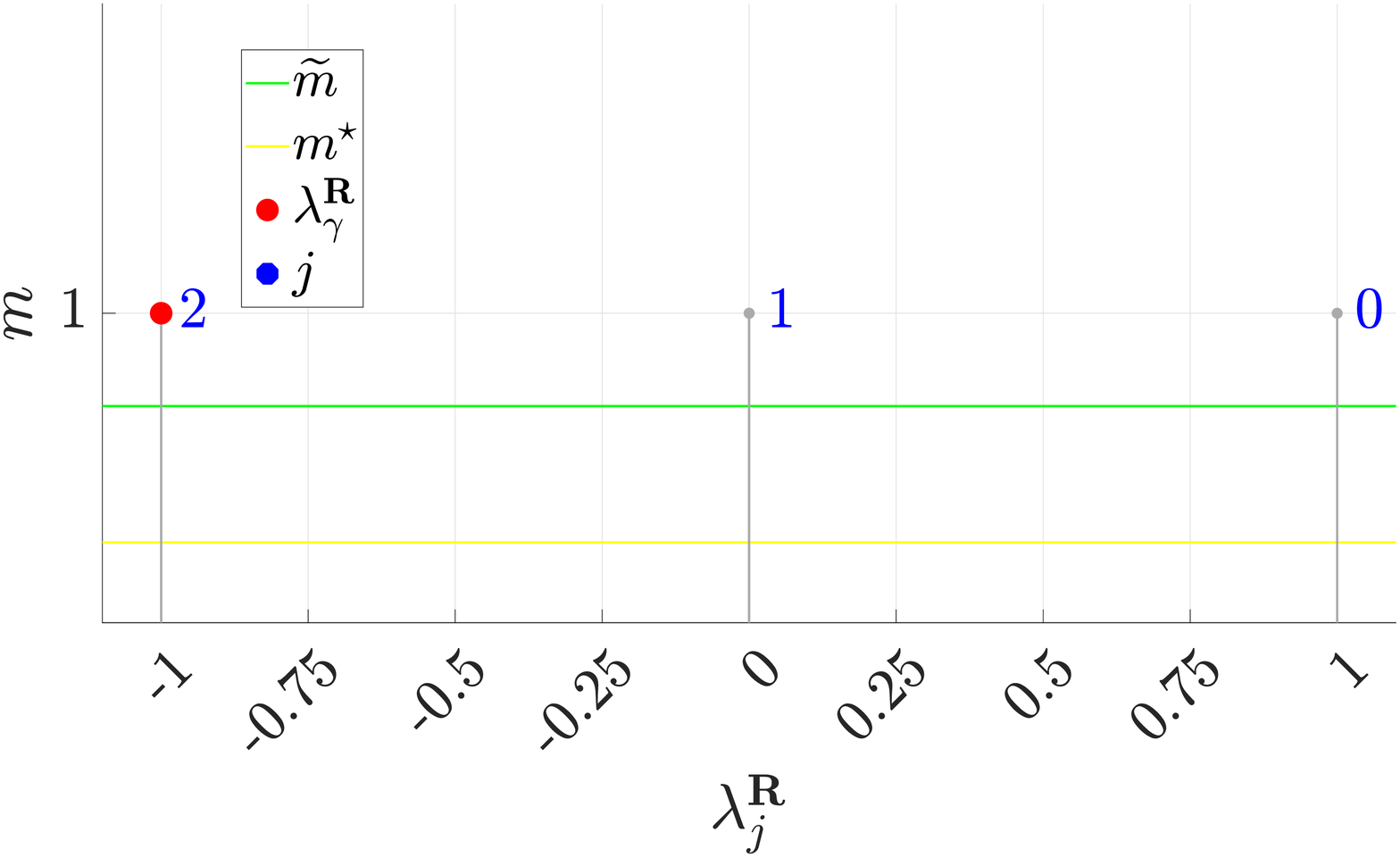}\label{fig:n4}}\hspace{0.1cm}
	\subfigure[$N=5$]{\includegraphics[width=0.49\columnwidth]{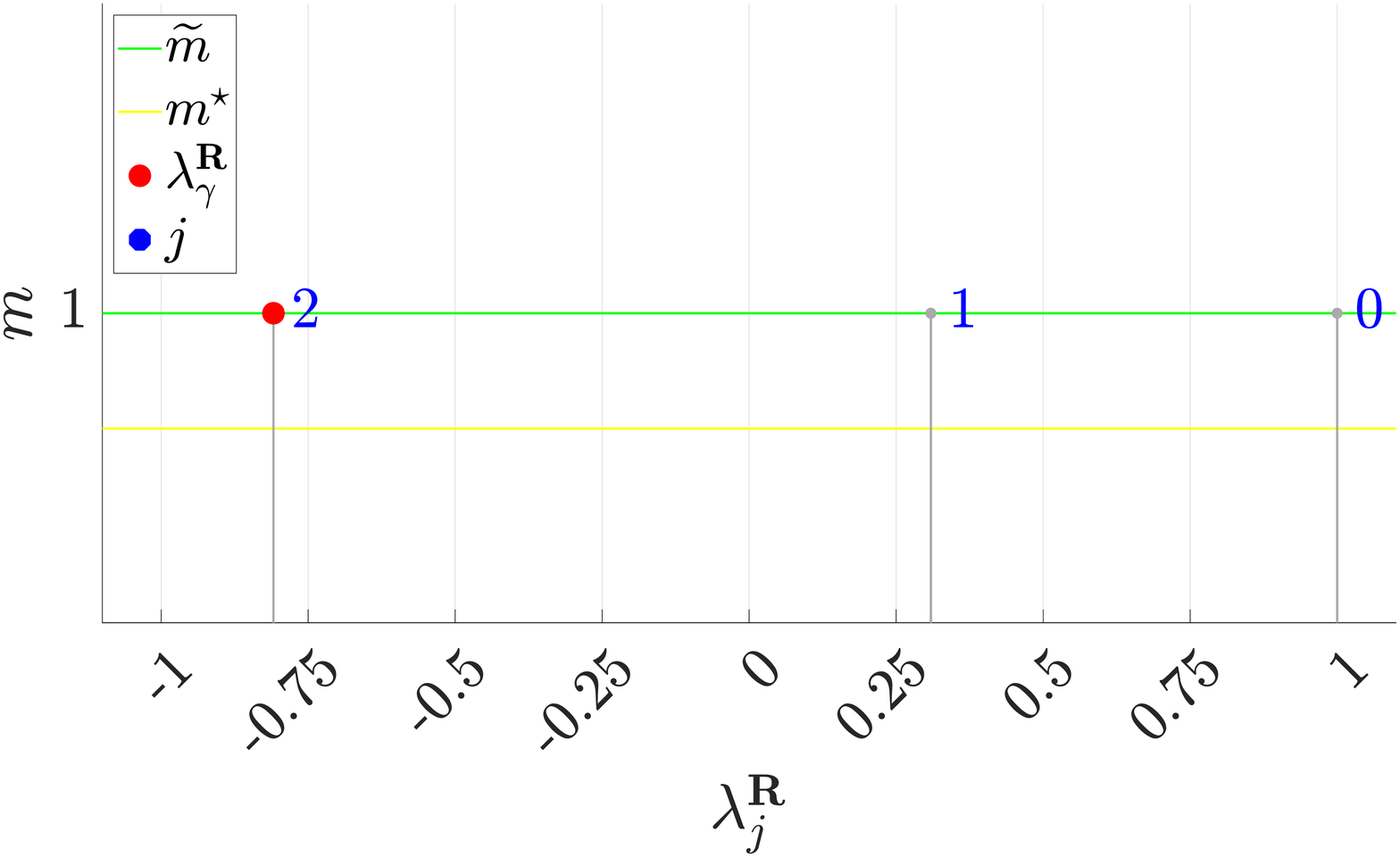}\label{fig:n5}}\vspace{-0.2cm}\\
	\subfigure[$N=6$]{\includegraphics[width=0.49\columnwidth]{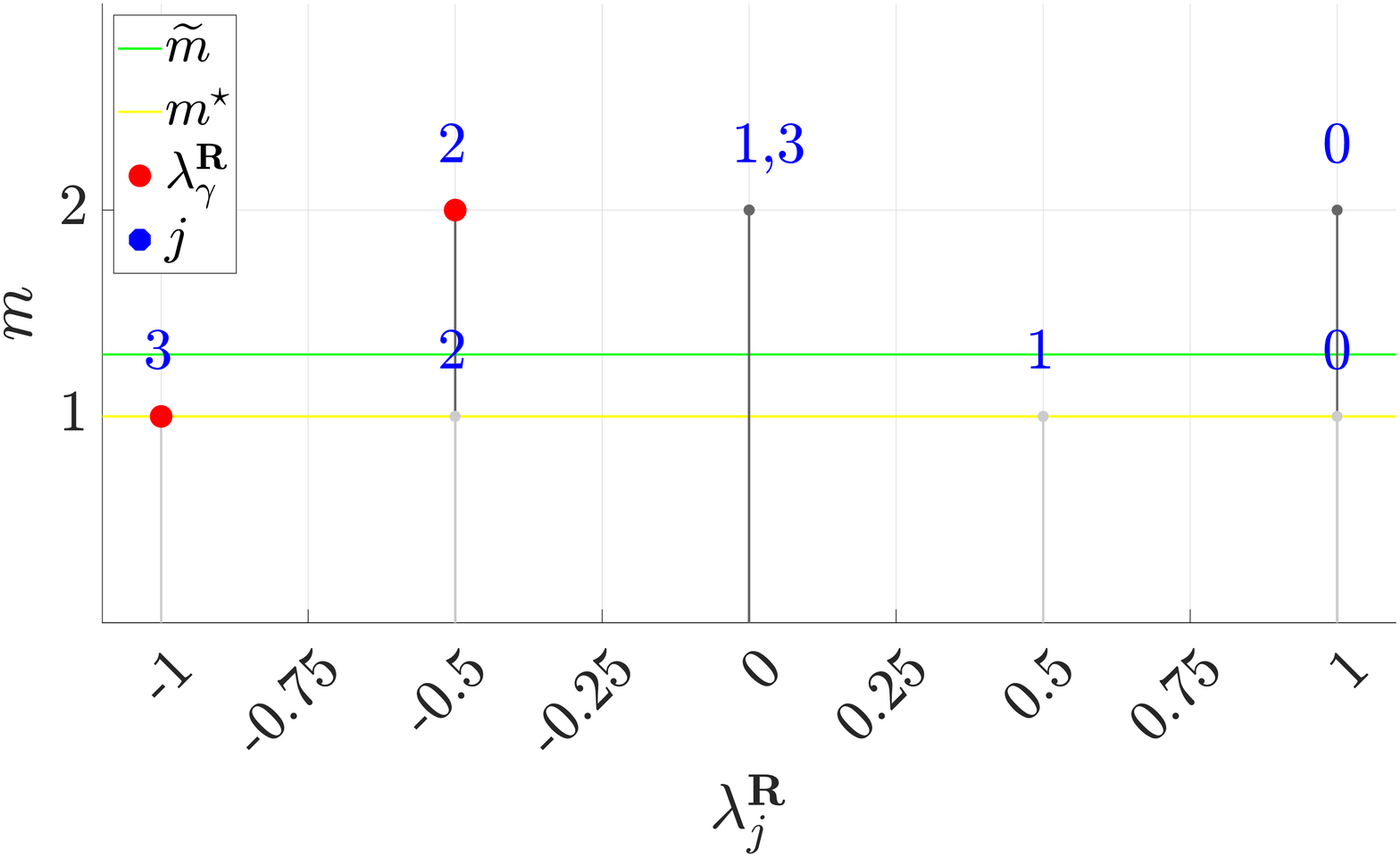}\label{fig:n6}}\hspace{0.1cm}
	\subfigure[$N=7$]{\includegraphics[width=0.49\columnwidth]{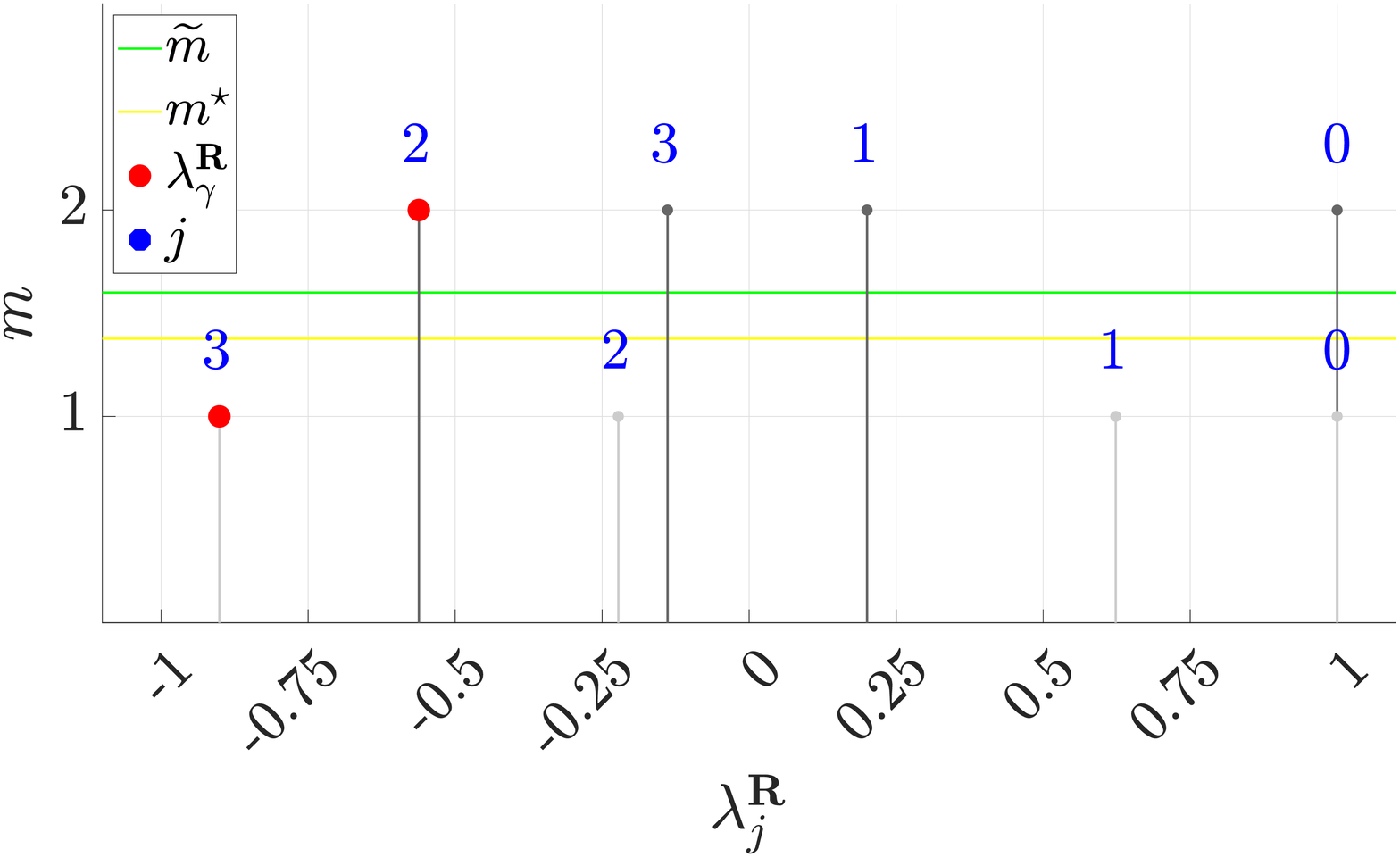}\label{fig:n7}}\vspace{-0.2cm}\\
	\subfigure[$N=8$]{\includegraphics[width=0.49\columnwidth]{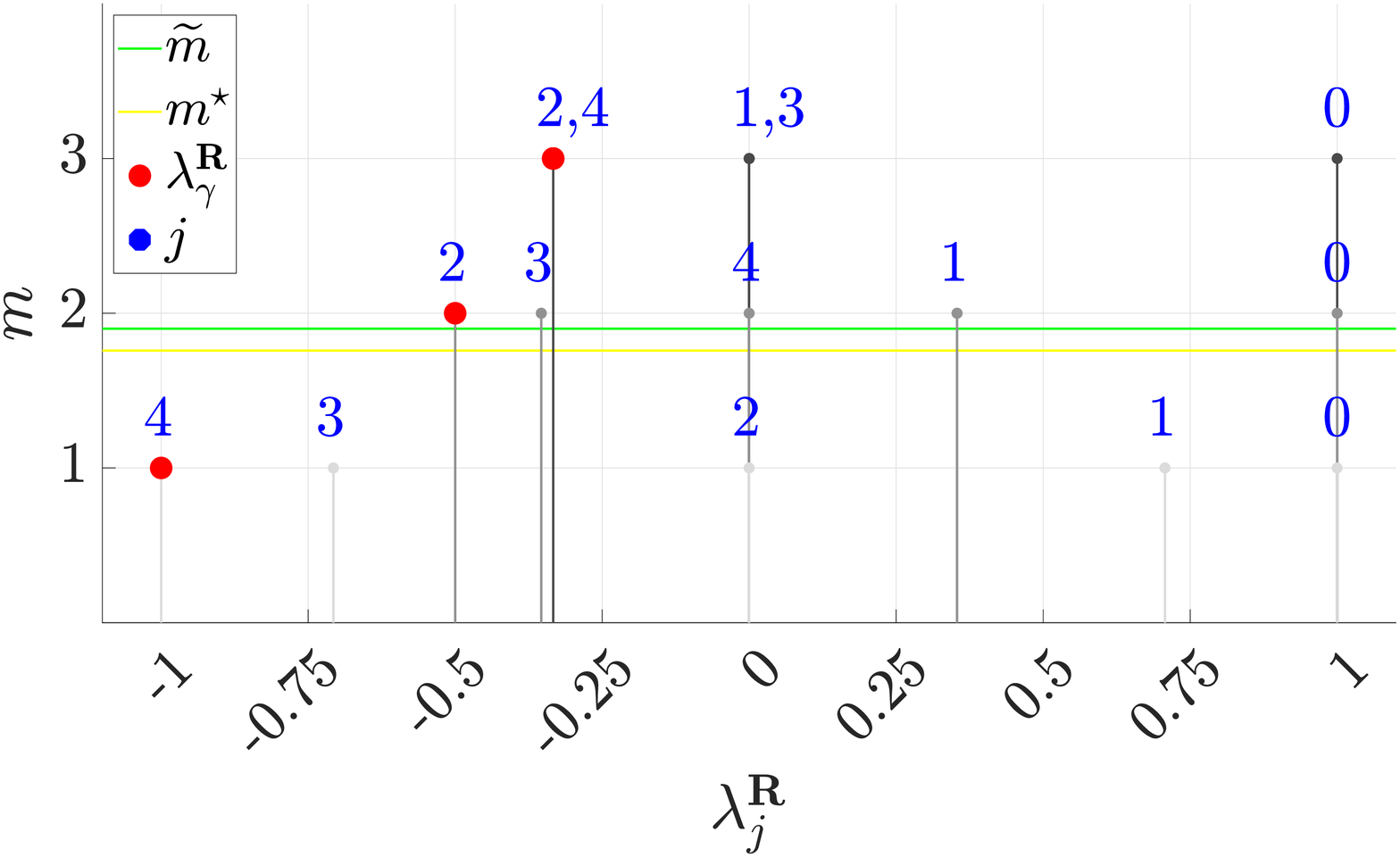}\label{fig:n8}}\hspace{0.1cm}
	\subfigure[$N=9$]{\includegraphics[width=0.49\columnwidth]{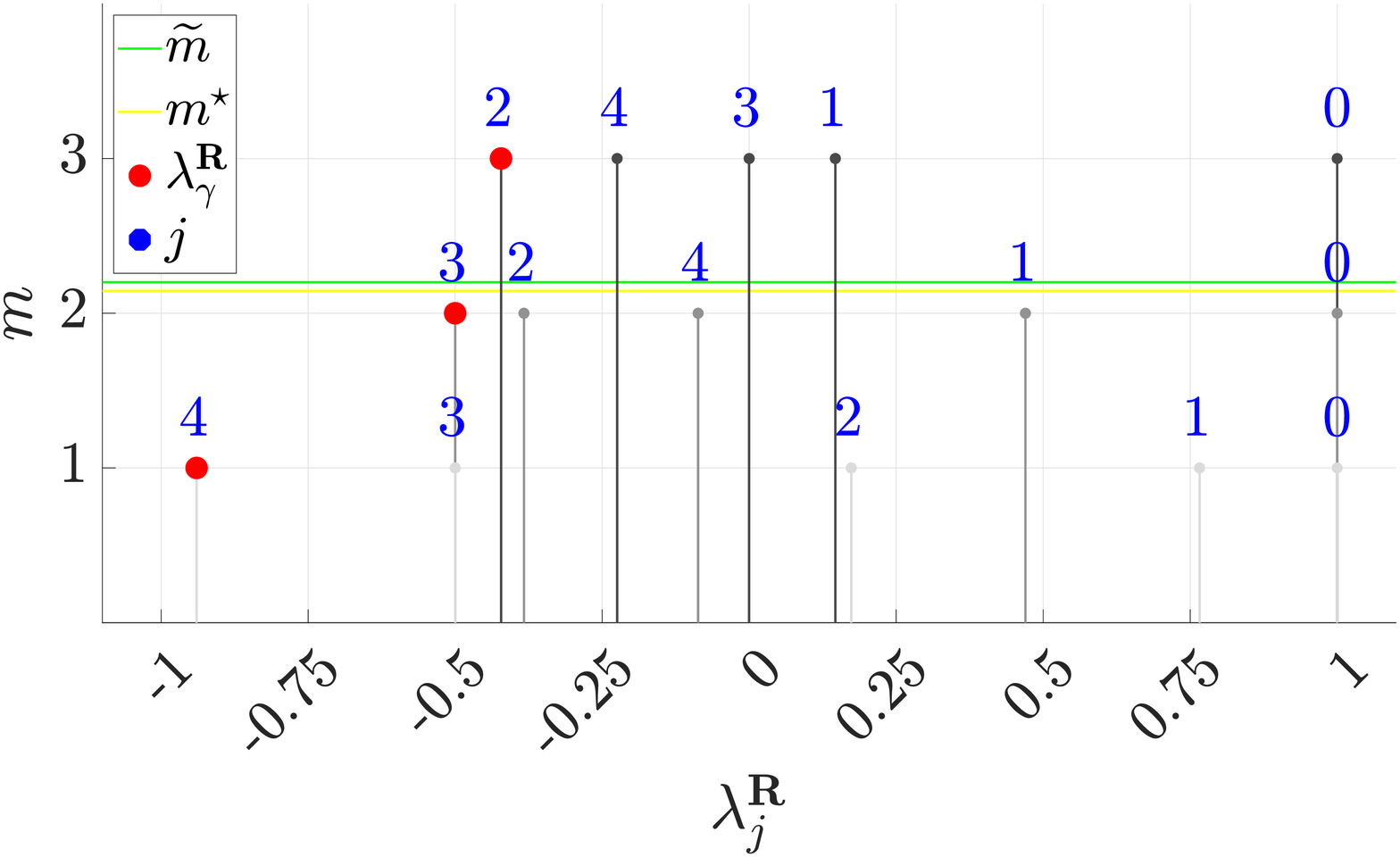}\label{fig:n9}}\vspace{-0.2cm}\\
	\subfigure[$N=10$]{\includegraphics[width=0.49\columnwidth]{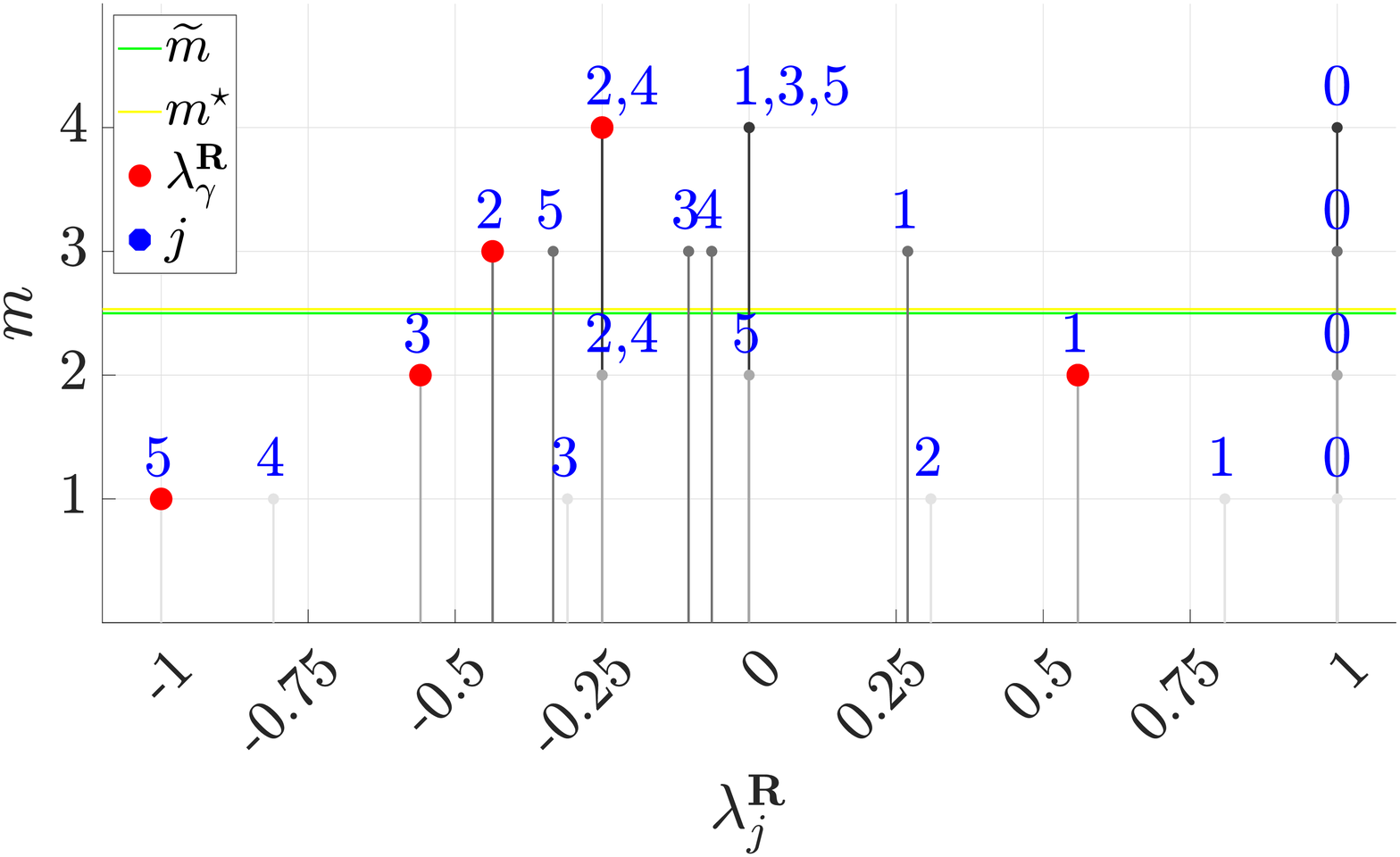}\label{fig:n10}}\hspace{0.1cm}
	\subfigure[$N=11$]{\includegraphics[width=0.49\columnwidth]{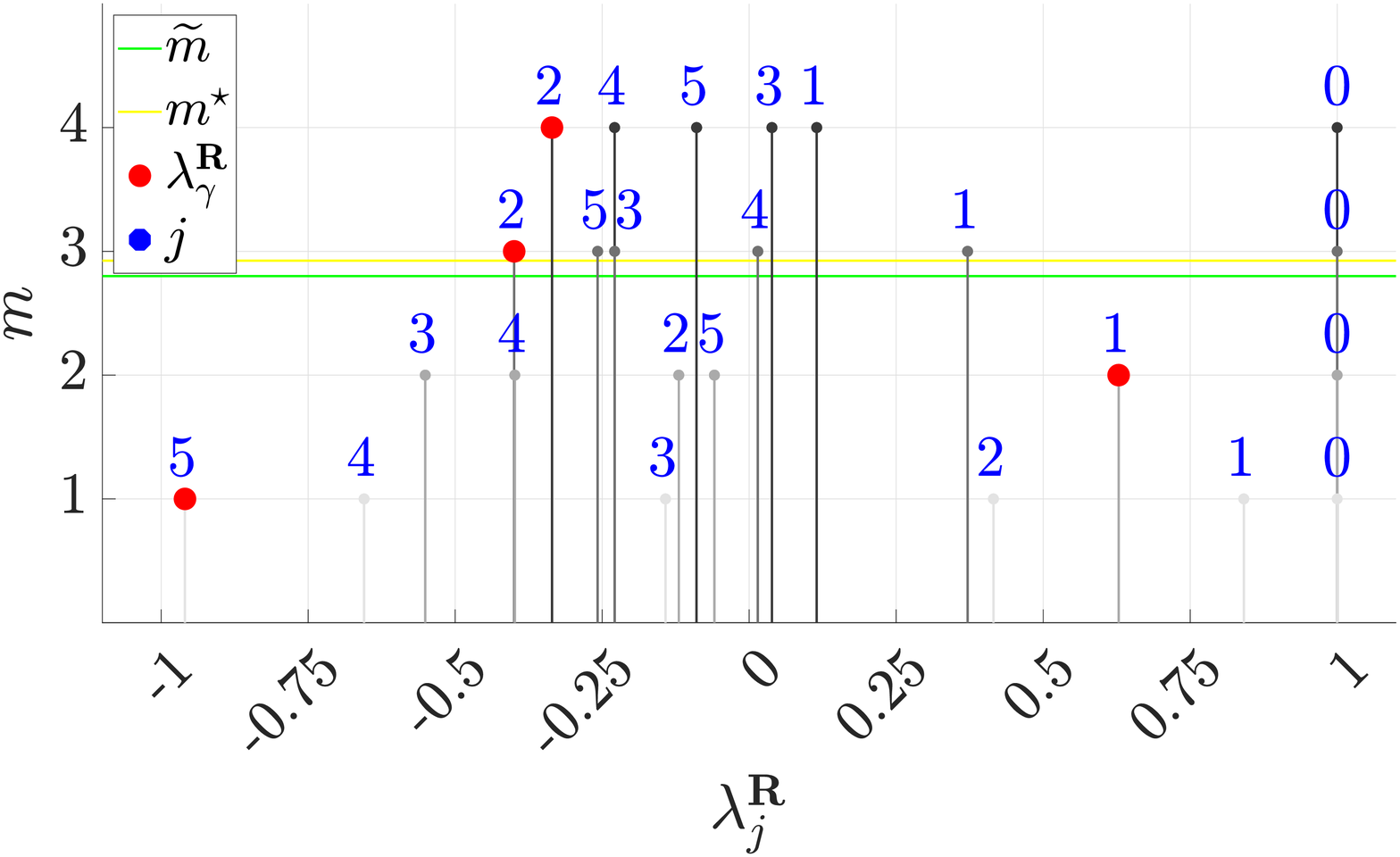}\label{fig:n11}}\vspace{-0.2cm}
	\caption{General eigenvalue distribution of the \Randic matrix spectrum $\Lambda(\R)$ for the RRLs $C_{N}^{m}$ with $N=4,\ldots, 11$ and $m = 1,\ldots ,n-1 = \lfloor N/2\rfloor-1$.}
	\label{fig:eig_distr}		
\end{figure}

A few observations made on the pattern of values taken by $\sigma(\R)$ are here provided. In this direction, examples in Fig. \ref{fig:eig_distr} grant to cover some of the most important aspects of this research, depicting a graphical representation of the spectrum $\Lambda(\R)$. Specifically, each diagram in Fig. \ref{fig:eig_distr} shows how the eigenvalues $\lambda_{j}^{\R}$ spread over the interval $[-1,1]$, as the order $m$ changes for a fixed size $N$, with $4\leq N \leq 11$. 
Plots \ref{fig:n4}-\ref{fig:n11} also illustrate in blue all indexes $j = 0,\ldots,n$ for relation \eqref{eq:lambdaRj}, thresholds $m^{\star}$ and $\widetilde{m}$ (see point \ref{point:mstarm2} in Rmk. \ref{rmk:ifconj1true}) with a yellow and a green line respectively, and the eigenvalue $\lambda^{\R}_{\gamma}$ with a red dot (where $\gamma$ is defined as in \eqref{eq:characterizationofjcirc}). 

With regard to Fig. \ref{fig:eig_distr}, it is possible to observe the following facts descending from all the previous statements presented in Sec. \ref{sec:main_results}.

\begin{itemize}
\item $\lambda^{\R}_{j} \in [-1,1]$ holds $\forall j\in \Jsh$, with $-1$ and $1$ simple eigenvalues.
\item $\lambda_{1}^{\R} = \lambda_{N-1}^{\R} > \lambda_{j}^{\R}$ holds for all $j\in \{2,\ldots,n\}$.
\item For $m=1$, one has $\lambda_{\gamma}^{\R}$ with $\gamma =n$; and if $N$ is even then $\lambda_{\gamma}^{\R} = -1$.
\item If $m\geq m^{\star}$ then $\lambda_{j^{\star}}^{\R} = \lambda_{2}^{\R} = \lambda_{N-2}^{\R}$ and if $m \geq \max\{m^{\star},\widetilde{m}\}$ then $\lambda_{\gamma}^{\R} = \lambda_{2}^{\R} = \lambda_{N-2}^{\R}$, thus supporting property \ref{point:mstarm2} in Rmk. \ref{rmk:ifconj1true}.
\end{itemize}

\clearpage

To provide further evidences to the speculations made in Rmk. \ref{rmk:ifconj1true}, some peculiarities and patterns can be also found for the following values of $N$.

\begin{itemize}
\item For $N=5$ one has $\widetilde{m}=1$ and, consequently, property \ref{point:m2l2} in Rmk. \ref{rmk:ifconj1true} holds tightly.
\item For $N=6$ one has $m^{\star}=1$; hance, if $m=2 > m^{\star}$, the information about $\widetilde{m}$ becomes necessary in order to satisfy property \ref{point:mstarm2} in Rmk. \ref{rmk:ifconj1true}.
\item For $N=10$ and $m=2$ one has $\sigma(\R)= \sqrt{5}/4 = \lambda_{1}^{\R} = \lambda_{9}^{\R} = -\lambda_{3}^{\R} = -\lambda_{7}^{\R}$, i.e. $\gamma$ takes both the values in $\{1,n\}$. Moreover, in this case, it holds that $m^{\star} \approx 2.5330 > 2.5 = \widetilde{m} $, conversely to the previous cases with $N=5$ and $N=6$. 
\end{itemize}

To sum up, each debated example in Fig. \ref{fig:eig_distr} gravitates, to some extent, around the key relation in \eqref{eq:lambdaRj}, describing the spectrum $\Lambda(\R)$ of the \Randic matrix. It is important to recall that this investigation completely leverages the fundamental idea of studying the spectral properties of RRLs via the Dirichlet kernel redefined as in \eqref{eq:Diri_real_expr}. Further clues are also given to support claims in Ssec. \ref{ssec:conjubstar}.

\subsection{Conjecture on the values taken by $\sigma(\R)$}

All the previous discussions suggest few clues about the possibility of computing exactly $\sigma(\R)$ by understanding the behavior of index $\gamma$ defined in \eqref{eq:characterizationofjcirc}. The exact knowledge of the essential spectral radius of $\R$ is also motivated by various research areas, such as the convergence analysis of Page Rank and random walk processes \cite{ChungZhao2010}. 

Remarkably, from the numerical examples given in Ssec. \ref{Ssec:numerical_examples}, it is possible to observe the following facts. Graph $C_{9}^{2}$ in Fig. \ref{fig:n9} is the unique example leading to $\gamma=3$ only (if $m\geq 2$), as $\sigma(\R) = -\lambda_{3}^{\R} = -1/2 > \lambda_{1}^{\R} \approx 0.4698$. Graph $C_{10}^{2}$ in Fig. \ref{fig:n10} is the unique example leading to both $\gamma=1$ and $\gamma=3$, as $\sigma(\R) = \lambda_{1}^{\R}  = -\lambda_{3}^{\R} = \sqrt{5}/4$. In each diagram of Fig. \ref{fig:eig_distr} it holds that $\gamma = n$, if and only if $m=1$, or $\gamma = 2$, if and only if $m\geq \max\{m^{\star},\widetilde{m}\}$. In the remaining cases, it holds that $\gamma=1$. Therefore, the following conjecture is drawn after having run some numerical tests\footnote{These are performed for all $N$ and $m$ such that $4 \leq N \leq 10000$ and $1 \leq m < n$.}.

\begin{conj}[Characterization of the essential spectral radius index $\gamma$]\label{conj:exactsigmaR}
Let $m^{\star}$ and $\widetilde{m} $ be defined as in \eqref{eq:mstar} and \eqref{eq:tildem}, respectively.
For all $N\geq 4$, the essential spectral radius $\sigma(\R) = \vert \lambda_{\gamma}^{\R}\vert = \vert \lambda_{N-\gamma}^{\R}\vert $ associated to the \Randic matrix $\R$ of a RRL $C_{N}^{m}$ can be computed through index
\begin{equation}\label{eq:gammaindexing}
\gamma = \begin{cases}
	n, \quad &\text{if } N \geq 8 \text{ and } m = 1; \\
	3, \quad &\text{if } N = 6,7 \text{ and } m = 1 \text{ or if } N = 9,10 \text{ and } m = 2; \\
	2, \quad &\text{if } N \geq 4 \text{ and } m \geq \min\{n-1,\max\{m^{\star},\widetilde{m}\}\}; \\
	1, \quad &\text{otherwise}.
\end{cases}
\end{equation}
Furthermore, a complete characterization of $\gamma$ is given by taking into account \eqref{eq:gammaindexing} along with the fact that $\gamma = 1$ also holds in the following four cases: (i) for all odd $N\geq 5$ and $m=1$; (ii) for all $N\geq 4$ and $m=\max\{m^{\star},\widetilde{m}\}$; (iii) for $N=10$ and $m=2$; (iv) for all even $N\geq 4$ and $m=n-1$.
\end{conj}


\section{Conclusions and future directions}\label{sec:concl_future_dirs}
In this work, a peculiar class of circulant graphs, referred to as \textit{regular ring lattices}, is described highlighting the relationship between the spectrum of their characteristic matrices and the well-known Dirichlet kernel. Several properties related to the eigenvalues are described extensively, with a particular focus on the Fiedler value, the spectral radius of the Laplacian and the essential spectral radius of the \Randic matrix associated to these graphs. Part of the proven results is also discussed in details with auxiliary diagrams depicting the related spectral distributions. Furthermore, conjectures on the computation of the debated spectral quantities are formulated and their formal demonstration is envisaged.


\section*{Acknowledgements}
A special thank goes to my Ph.D. advisor prof. Angelo Cenedese, who encouraged and supported me during this study.

\bibliography{biblio}

\begin{thebibliography}{10}
\expandafter\ifx\csname url\endcsname\relax
  \def\url#1{\texttt{#1}}\fi
\expandafter\ifx\csname urlprefix\endcsname\relax\def\urlprefix{URL }\fi
\expandafter\ifx\csname href\endcsname\relax
  \def\href#1#2{#2} \def\path#1{#1}\fi

\bibitem{BarahonaPecora2002}
M.~Barahona, L.~M. Pecora, Synchronization in small-world systems, Physical
  review letters 89~(5) (2002) 054101.

\bibitem{WuBarahonaTan2011}
J.~Wu, M.~Barahona, Y.-J. Tan, H.-Z. Deng, Robustness of regular ring lattices
  based on natural connectivity, International Journal of Systems Science
  42~(7) (2011) 1085--1092.

\bibitem{McKeeArumugam2013}
T.~A. McKee, S.~Arumugam, Graphs that induce only k-cycles, AKCE International
  Journal of Graphs and Combinatorics 10~(1) (2013) 29--36.

\bibitem{HelleSimonyi2016}
Z.~Helle, G.~Simonyi, Orientations making k-cycles cyclic, Graphs \&
  Combinatorics 32~(6) (2016) 2415--2423.

\bibitem{Gray2005}
R.~M. Gray, Toeplitz and Circulant Matrices: A Review, no. 2:3, Now -
  Foundations and Trends in Communications and Information Theory, Boston -
  Delf, MA, US, 2005, pag. 34.

\bibitem{MakhdoumiOzdaglar2017}
A.~{Makhdoumi}, A.~{Ozdaglar}, Convergence rate of distributed admm over
  networks, IEEE Transactions on Automatic Control 62~(10) (2017) 5082--5095.

\bibitem{Spielman2007}
D.~A. {Spielman}, Spectral graph theory and its applications, in: 48th Annual
  IEEE Symposium on Foundations of Computer Science (FOCS'07), 2007, pp.
  29--38.

\bibitem{LovisariZampieri2012}
E.~Lovisari, S.~Zampieri, Performance metrics in the average consensus problem:
  A tutorial, Annual Reviews in Control 36~(1) (2012) 26 -- 41.

\bibitem{FranceschelliGasparriGiua2013}
M.~Franceschelli, A.~Gasparri, A.~Giua, C.~Seatzu, Decentralized estimation of
  laplacian eigenvalues in multi-agent systems, Automatica 49~(4) (2013) 1031
  -- 1036.

\bibitem{FabrisMichielettoCenedese2019}
M.~Fabris, G.~Michieletto, A.~Cenedese, On the distributed estimation from
  relative measurements: a graph-based convergence analysis, in: 2019 18th
  European Control Conference (ECC), 2019, pp. 1550--1555.

\bibitem{FabrisMichielettoCenedese2020}
M.~Fabris, G.~Michieletto, A.~Cenedese, A proximal point approach for
  distributed system state estimation, IFAC-PapersOnLine 53~(2) (2020)
  2702--2707, 21st IFAC World Congress.

\bibitem{FabrisMichielettoCenedese2022}
M.~Fabris, G.~Michieletto, A.~Cenedese, A general regularized distributed
  solution for system state estimation from relative measurements, IEEE Control
  Systems Letters 6 (2022) 1580--1585.

\bibitem{LiuLiuMuhammad2018}
G.~{Liu}, S.~{Liu}, K.~{Muhammad}, A.~K. {Sangaiah}, F.~{Doctor}, Object
  tracking in vary lighting conditions for fog based intelligent surveillance
  of public spaces, IEEE Access 6 (2018) 29283--29296.

\bibitem{HenriquesCaseiroMartins2012}
J.~F. Henriques, R.~Caseiro, P.~Martins, J.~Batista, Exploiting the circulant
  structure of tracking-by-detection with kernels, in: A.~Fitzgibbon,
  S.~Lazebnik, P.~Perona, Y.~Sato, C.~Schmid (Eds.), Computer Vision -- ECCV
  2012, Springer Berlin Heidelberg, Berlin, Heidelberg, 2012, pp. 702--715.

\bibitem{AlpagoZorziFerrante2018}
D.~{Alpago}, M.~{Zorzi}, A.~{Ferrante}, Identification of sparse reciprocal
  graphical models, IEEE Control Systems Letters 2~(4) (2018) 659--664.

\bibitem{OrtegaFrossardKovacevic2018}
A.~{Ortega}, P.~{Frossard}, J.~{Kova\u{c}evi\'{c}}, J.~M.~F. {Moura},
  P.~{Vandergheynst}, Graph signal processing: Overview, challenges, and
  applications, Proceedings of the IEEE 106~(5) (2018) 808--828.

\bibitem{SevesoBenedettiParis2019}
L.~Seveso, C.~Benedetti, M.~G.~A. Paris, The walker speaks its graph: global
  and nearly-local probing of the tunnelling amplitude in continuous-time
  quantum walks, Journal of Physics A: Mathematical and Theoretical 52~(10)
  (2019) 105304.

\bibitem{ShuAhuja2011}
X.~{Shu}, N.~{Ahuja}, Imaging via three-dimensional compressive sampling
  (3dcs), in: 2011 International Conference on Computer Vision, 2011, pp.
  439--446.

\bibitem{AntholzerWolfSandbichler2019}
S.~Antholzer, C.~Wolf, M.~Sandbichler, M.~Dielacher, M.~Haltmeier, Compressive
  time-of-flight 3d imaging using block-structured sensing matrices, Inverse
  Problems 35~(4) (2019) 045004.

\bibitem{GastparVetterli2005}
M.~{Gastpar}, M.~{Vetterli}, Power, spatio-temporal bandwidth, and distortion
  in large sensor networks, IEEE Journal on Selected Areas in Communications
  23~(4) (2005) 745--754.

\bibitem{GancioRubido2022}
J.~Gancio, N.~Rubido, Critical parameters of the synchronisation's stability
  for coupled maps in regular graphs, Chaos, Solitons \& Fractals 158 (2022)
  112001.

\bibitem{Fiedler1973}
M.~Fiedler, Algebraic connectivity of graphs, Czechoslovak Mathematical Journal
  23~(2) (1973) 298--305.

\bibitem{Cheeger1969}
J.~Cheeger, A lower bound for the smallest eigenvalue of the laplacian, in:
  Proceedings of the Princeton conference in honor of Professor S. Bochner,
  1969, pp. 195--199.

\bibitem{LiGuoShiu2013}
J.~Li, J.-M. Guo, W.~C. Shiu, On the second largest laplacian eigenvalues of
  graphs, Linear Algebra and its Applications 438~(5) (2013) 2438 -- 2446.

\bibitem{HuiqingMeiFeng2004}
H.~Liu, M.~Lu, F.~Tian, On the laplacian spectral radius of a graph, Linear
  Algebra and its Applications 376 (2004) 135--141.

\bibitem{Shi2007}
L.~Shi, Bounds on the (laplacian) spectral radius of graphs, Linear Algebra and
  its Applications 422~(2) (2007) 755--770.

\bibitem{CarliFagnaniSperanzon2008}
R.~Carli, F.~Fagnani, A.~Speranzon, S.~Zampieri, Communication constraints in
  the average consensus problem, Automatica 44~(3) (2008) 671--684.

\bibitem{BrinonSchenato2013}
L.~Bri\~{n}\'{o}n Arranz, L.~Schenato, Consensus-based source-seeking with a
  circular formation of agents, in: 2013 European Control Conference (ECC),
  2013, pp. 2831--2836.

\bibitem{chung1997spectral}
F.~R. Chung, F.~C. Graham, Spectral graph theory, no.~92, American Mathematical
  Society, Providence, RI, US, 1997.

\bibitem{ZhangDong2011}
X.-D. {Zhang}, {The Laplacian eigenvalues of graphs: a survey}, arXiv e-prints
  (2011) arXiv:1111.2897\href {http://arxiv.org/abs/1111.2897}
  {\path{arXiv:1111.2897}}.

\bibitem{KleinRandic1993}
D.~J. Klein, M.~Randi{\'c}, Resistance distance, Journal of mathematical
  chemistry 12~(1) (1993) 81--95.

\bibitem{BanerjeeMehatari2016}
A.~Banerjee, R.~Mehatari, An eigenvalue localization theorem for stochastic
  matrices and its application to randi\'{c} matrices, Linear Algebra and its
  Applications 505 (2016) 85 -- 96.

\bibitem{Rojo2007}
O.~Rojo, A nontrivial upper bound on the largest laplacian eigenvalue of
  weighted graphs, Linear Algebra and its Applications 420~(2) (2007) 625 --
  633.

\bibitem{Sorgun2013}
S.~Sorgun, Bounds for the largest laplacian eigenvalue of weighted graphs,
  International Journal of Combinatorics 2013 (2013) 1--8.

\bibitem{Euler1741}
L.~Euler, Solutio problematis ad geometriam situs pertinentis, Commentarii
  academiae scientiarum Petropolitanae (1741) 128--140.

\bibitem{BruncknerBruncknerThomson1997}
A.~M. Brunckner, J.~B. Brunckner, B.~S. Thomson, Real Analysis, Pearson
  Prentice Hall, Upper Saddle River, NJ, US, 1997.

\bibitem{Wiggins2007}
A.~Wiggins, The minimum of the dirichlet kernel, notes. Webpage:
  \url{www-personal.umd.umich.edu/~adwiggin/TeachingFiles/FourierSeries/Resources/DirichletKernel.pdf}
  (2007).

\bibitem{Kirkwood2018}
J.~Kirkwood, Mathematical physics with partial differential equations, 2nd
  Edition, Academic Press, 125 London Wall, London EC2Y 5AS, United Kingdom,
  2018.

\bibitem{PetersenPedersen2008}
K.~B. Petersen, M.~S. Pedersen, et~al., The matrix cookbook, Technical
  University of Denmark 7~(15) (2008) 510.

\bibitem{landau1981bounds}
H.~Landau, A.~Odlyzko, Bounds for eigenvalues of certain stochastic matrices,
  Linear algebra and its Applications 38 (1981) 5--15.

\bibitem{AbramowitzStegun1972}
M.~Abramowitz, I.~A. Stegun, Handbook of Mathematical Functions With Formulas,
  Graphs, and Mathematical Tables, no.~55, National Bureau of Standards Applied
  Mathematics Series, Washington D.C., US, 1972, pag. 17,75.

\bibitem{LiuLu2010}
H.~Liu, M.~Lu, Bounds for the laplacian spectral radius of graphs, Linear and
  Multilinear Algebra 58~(1) (2010) 113--119.

\bibitem{GarinSchenato2010}
F.~Garin, L.~Schenato, A Survey on Distributed Estimation and Control
  Applications Using Linear Consensus Algorithms, Springer London, London,
  2010, pp. 75--107.

\bibitem{ChungZhao2010}
F.~Chung, W.~Zhao, PageRank and Random Walks on Graphs, Springer Berlin
  Heidelberg, Berlin, Heidelberg, 2010, pp. 43--62.

\end{thebibliography}

\end{document}